\documentclass[smallextended]{svjour3}
\smartqed

\usepackage{graphicx}
\usepackage{amsmath}
\usepackage{amssymb}
\usepackage{hyperref}
\usepackage{booktabs}
\usepackage{subcaption}
\usepackage{float}
\usepackage{multirow}
\usepackage[numbers, sort&compress]{natbib}
\begin{document}

\title{A GRMHD-Calibrated Semi-Analytical Model for Hot Sub-Keplerian Accretion Flows in Kerr Spacetime}


\author{
Nabin Bhusal \and
Manil Khatiwada \and
Yogesh Maharjan \and
David Garofalo \and
Chandra B. Singh
}


\institute{
Nabin Bhusal \at
Central Department of Physics, Tribhuvan University,
Kirtipur, Kathmandu 44600, Nepal \\
\email{nabinbhusal25@gmail.com}
\and
Manil Khatiwada \at
Central Department of Physics, Tribhuvan University,
Kirtipur, Kathmandu 44600, Nepal
\and
Yogesh Maharjan \at
Faculty of Physics and Applied Informatics,
University of Łódź,
Pomorska 149/153,
90-236 Łódź,
Poland
\and
David Garofalo \at
Department of Physics,
Kennesaw State University,
Marietta, GA 30060,
USA
\and
Chandra B. Singh \at
Calle 134A No. 148A 21,
San Pedro de Tibabuyes,
Suba,
Bogotá,
Colombia \\
\email{chandratalk@gmail.com}
}

\date{Received: date / Accepted: date}

\maketitle

	\begin{abstract}
        
        We develop a simple, semi-analytical, kinematic model for hot, thick accretion flows, constructed by interpolating between Keplerian and free-fall geodesic solutions in the Kerr metric. Unlike self-consistent general relativistic magnetohydrodynamics (GRMHD) frameworks, our model contains no explicit magnetic fields or stress terms; instead, it uses a smooth, radially varying transition function \(T(r)\) to connect the velocity components from near-Keplerian rotation at large distances to a free-fall state near the event horizon. While the coefficients $\alpha$ and $\beta$ remain constant, the transition function is a true function of radius, allowing the flow properties to vary smoothly with radius. We calibrate and validate this model against time- and azimuthally averaged profiles from long-duration magnetically arrested disk (MAD) simulations spanning a wide range of black hole spins ($a=-0.9$ to $+0.9$). The model successfully captures the properties of accretion flow parameters across both prograde and retrograde configurations. Quantitatively, the predicted radial velocity, angular velocity, and density profiles match the simulation data to within an average factor of approximately 1.8, 1.6, and 1.6, respectively, while the specific angular momentum exhibits the closest agreement, remaining within a factor of approximately 1.2. This fast semi-analytical prescription gives significantly lower errors than previous constant-coefficient models and it is a computationally affordable tool for various applications such as ray tracing, accretion parameter exploration, and spectral modelling.

\end{abstract}

\section{Introduction}
\label{intro}

Accretion onto black holes is a fundamental process in high-energy astrophysics that powers a wide range of astronomical phenomena, including X-ray binaries and active galactic nuclei (AGN) \cite{Frank2002}. In these systems, gravitational potential energy is efficiently converted into radiation as matter spirals toward the event horizon \cite{kato2008}. The physics of accretion becomes especially complex in the strong-field regime, where general relativistic effects play a crucial role in governing the dynamics of the flow. For rotating black holes, the surrounding spacetime is described by the Kerr metric \cite{kerr1963}, an exact axisymmetric solution of Einstein’s field equations that determines particle motion, angular momentum transport, and energy dissipation within the accretion disk in geometric units $(G=c=M=1)$ \cite{boyer1967maximal}:
\begin{equation}
ds^2 = - \left(1-\frac{2r}{\Sigma}\right) dt^2 - \frac{4ar \sin^2\theta}{\Sigma} dt\,d\phi 
+ \frac{\Sigma}{\Delta} dr^2 + \Sigma d\theta^2 + \frac{\mathcal{A} \sin^2\theta}{\Sigma} d\phi^2,
\end{equation}
The metric functions are defined in terms of the radial coordinate $r$, the polar angle $\theta$ and the dimensionless black hole spin parameter $a$ \cite{rezzolla2013relativistic}. Here, $\Sigma = r^{2} + a^{2}\cos^{2}\theta$, $\Delta = r^{2} - 2r + a^{2}$, and $\mathcal{A} = (r^{2} + a^{2})^{2} - a^{2}\Delta \sin^{2}\theta$.

  For thin radiatively efficient accretion disks, the flow is typically confined near the equatorial plane ($\theta = \pi/2$), where the relevant dynamical variables are the radial four-velocity $u^r$ and the angular velocity \(\Omega = {d\phi}/{dt} = {u^\phi}/{u^t}\) and the radial component of the four-velocity is much smaller than the azimuthal component. Therefore, matter approximately follows circular geodesics \cite{NovikovThorne1973}. The Keplerian angular velocity for circular orbits in the equatorial plane is obtained from the condition $dV_{\rm eff}/dr = 0$, where $V_{\rm eff}$ is the effective potential \cite{bardeen1972}. The corresponding radial velocity vanishes in the ideal Keplerian disk, \(u^r_K = 0,\) indicating full centrifugal support. Under certain perturbations, Keplerian disks may also develop eccentric modes and related instabilities \cite{adams1989eccentric}. This describes the outer disk regions, where angular momentum transport occurs gradually via viscosity or magnetic stresses \cite{papaloizou1995theory}. Closer to the black hole, inside the innermost stable circular orbit (ISCO), centrifugal support fails and matter plunges rapidly toward the horizon \cite{NovikovThorne1973, page1974disk}.

 Another class of accretion flows, radiatively inefficient accretion flows (RIAFs), is a geometrically thick and sub-Keplerian flow characterized by significant radial inflow and a two-temperature plasma \cite{Yuan2014, narayan1994advection}. Unlike thin disks, where centrifugal support dominates, RIAFs exhibit a substantial radial velocity component that can reach a significant fraction of the free-fall speed \cite{Yuan2014}. These properties make RIAFs particularly relevant for interpreting observations at millimeter and sub-millimeter wavelengths, especially those probing event-horizon-scale structures shown by the Event Horizon Telescope \cite{EHTC2022_I, Yuan2014}. This arises from the fact that RIAFs are optically thin, hot plasmas in which synchrotron emission dominates at millimeter and sub-millimeter wavelengths, enabling direct probing of the innermost accretion flow near the event horizon \cite{falcke2013toward, brinkerink2016asymmetric}. Despite their importance, many analytical and semi-analytical models of RIAFs rely on simplifying assumptions, typically formulated within the framework of Advection-Dominated Accretion Flows (ADAFs).  Early ADAF  solutions  assumed self-similarity and radially constant parameters, while more recent hybrid models \cite{Pu2016} combine Keplerian rotation and free-fall motion using constant coefficients, implying a fixed degree of rotational support. In addition to pressure-supported RIAFs, strongly magnetized accretion states known as Magnetically Arrested Disks (MADs) have emerged as an important framework for describing black hole accretion flows. GRMHD simulations show that MAD systems can produce geometrically thick and strongly magnetized flows, often associated with powerful relativistic jets. 

Realistic accretion flows are neither purely Keplerian nor purely in free-fall \cite{shakura2018accretion}. Accretion flows vary strongly with radius: they are nearly Keplerian at large radii and approach the free-fall regime near the black hole. To capture this behaviour, we adopt a prescription governed by the parameters $\alpha$ and $\beta$ that interpolates between the Keplerian and free-fall components \cite{Pu2016,Tiede2020}.
\begin{equation}
	u^r_{\rm sub} = u^r_K + (1-\alpha)\left(u^r_{\rm ff} - u^r_K\right),
	\label{eq:sub_no_blend1}
\end{equation}
\begin{equation}
	\Omega_{\rm sub} = \Omega_K + (1-\beta)\left(\Omega_{\rm ff} - \Omega_K\right),
	\label{eq:sub_no_blend2}
\end{equation}

where $\alpha$ and $\beta$ are constants determining the flow structure. By construction, $\alpha = 1$ and $\beta = 1$ correspond to the Keplerian regime, while $\alpha = 0$ and $\beta = 0$ correspond to the free-fall regime. The quantities with subscript ``ff'' correspond to the zero-angular-momentum free-fall geodesic, while ``K'' and ``sub'' denote the Keplerian and sub-Keplerian flow components, respectively. Here, $u^r$ is the radial four-velocity and $\Omega$ is the angular velocity. This hybrid approach has been successfully used to interpret emission from supermassive black holes on event-horizon scales, including Sgr A* and M87*. In this work, we revisit the model to show how rotational support gradually decreases as matter approaches the black hole while the inflow velocity smoothly increases. We introduce a radially varying transition function that governs the transition between Keplerian and free-fall dynamics. 

The objective of this work is not to replace GRMHD simulations or to reproduce the full MHD evolution of accretion flows. Rather, our goal is to develop a simple analytical approximation that captures the time-averaged radial structure of hot MAD accretion flows using only a small number of physically motivated parameters. Such a model can provide a practical alternative when full GRMHD calculations are computationally expensive.

This paper is organized as follows. Section~2 presents the theoretical framework, including the formulation of the semi-analytical model and the transition from nearly Keplerian to free-fall motion. Section~3 describes the model fitting procedure and calibration against the GRMHD simulation data. Section~4 presents the results and discusses the model performance through comparisons with the simulation profiles. Finally, Section~5 summarizes the main conclusions and outlines the limitations and future directions of this work.

\section{Governing Equations}

\subsection{Accretion Flow Parameters}

In Boyer-Lindquist coordinates, the normalization condition for the four-velocity, \(u^\mu u_\mu = -1,\) together with the assumption of equatorial motion ($u^\theta = 0$), gives
\begin{equation}
	(u^t)^2
	=
	\frac{1 + g_{rr}(u^r)^2}{K_0},
	\qquad
	K_0
	\equiv
	-\left(
	g_{tt}
	+
	\Omega^2 g_{\phi\phi}
	+
	2\Omega g_{t\phi}
	\right),
	\qquad
	\Omega
	\equiv
	\frac{u^\phi}{u^t}.
	\label{eq:ut_normalization}
\end{equation}

Thus, for a given angular velocity $\Omega(r)$ and radial four-velocity $u^r(r)$, Eq.~\eqref{eq:ut_normalization} uniquely determines $u^t$, while \(	u^\phi=\Omega u^t\),  and the specific angular momentum is defined as \(	\ell = - \frac{u_\phi}{u_t}\).
The nearly Keplerian and free-fall branches used to construct the blended accretion flow are summarized in Table~\ref{tab:keplerian_freefall}.

\begin{table}[ht]
	\centering
	\footnotesize
	\caption{Keplerian and free-fall branches used to construct the accretion flow model parameters (adapted from Pu, Akiyama \& Asada 2016).}
	\label{tab:keplerian_freefall}
	\renewcommand{\arraystretch}{1.4}
	\begin{tabular}{|l|c|c|}
		\hline
		\textbf{Branch} & $\mathbf{u^r}$ & $\mathbf{\Omega=u^\phi/u^t}$ \\
		\hline
		Keplerian ($r>r_0$)
		&
		$u^r_{\rm K}(r)=0$
		&
		$\displaystyle
		\Omega_{\rm K}(r)=\frac{1}{r^{3/2}+a}
		$
		\\
		\hline
		Keplerian ($r\le r_0$)
		&
		$\displaystyle
		u^r_{\rm K}(r)
		=
		-
		\left(\frac{2}{3r_0}\right)^{1/2}
		\left(\frac{r_0}{r}-1\right)^{3/2}
		$
		&
		$\displaystyle
		\Omega_{\rm K}(r)
		=
		\frac{\lambda+aH}
		{r^2+2r(1+H)}
		$
		\\
		\hline
		Free-fall
		&
		$\displaystyle
		u^r_{\rm ff}(r)
		=
		-
		\frac{\sqrt{2r(r^2+a^2)}}{\Sigma}
		$
		&
		$\displaystyle
		\Omega_{\rm ff}(r)
		=
		\frac{2ar}{A}
		$
		\\
		\hline
	\end{tabular}
\end{table}

The auxiliary quantities appearing in Table~\ref{tab:keplerian_freefall} are defined as
\begin{equation}
	\lambda=
	\frac{r_0^2-2a\sqrt{r_0}+a^2}
	{r_0^{3/2}-2\sqrt{r_0}+a},
	\qquad
	H=\frac{2r-a\lambda}{\Delta},
	\qquad
	r_0=r_{\rm ISCO}(a),
\end{equation}
where all quantities are evaluated on the equatorial plane ($\theta=\pi/2$), and $r_{\rm ISCO}(a)$ denotes the radius of the innermost stable circular orbit for a black hole with dimensionless spin parameter $a$. The free-fall branch describes a zero-angular-momentum geodesic released from rest at infinity, characterized by the conserved specific energy $E=-u_t=1$. In contrast, the ballistic branch for $r\le r_0$ conserves the specific angular momentum acquired at the ISCO, such that $\lambda\equiv\ell_{\rm ISCO}$, specific angular momentum at the ISCO.

For a steady, sub-Keplerian accretion flow, the quasi-spherical flow geometry is considered with a scale height satisfying $H/R \sim 1$ \cite{abramowicz1981rotation}. Under this assumption, mass conservation gives
\(	\dot{M}=4\pi r^{2}\rho_{\rm sub}(r)u_{\rm sub}^{r}(r),\)
where $\dot{M}$ is the mass accretion rate, $\rho_{\rm sub}(r)$ is the equatorial (midplane) density, and $u_{\rm sub}^{r}(r)$ is the sub-Keplerian radial four-velocity. 
\begin{equation}
	\rho_{\rm sub}(r)=
	\frac{\dot{M}}
	{4\pi r^{2}u_{\rm sub}^{r}(r)}.
	\label{eq:density_prescription}
\end{equation}

\subsection{Transition Function}

\begin{figure}[H]
	\centering
	\includegraphics[width=1.01\textwidth]{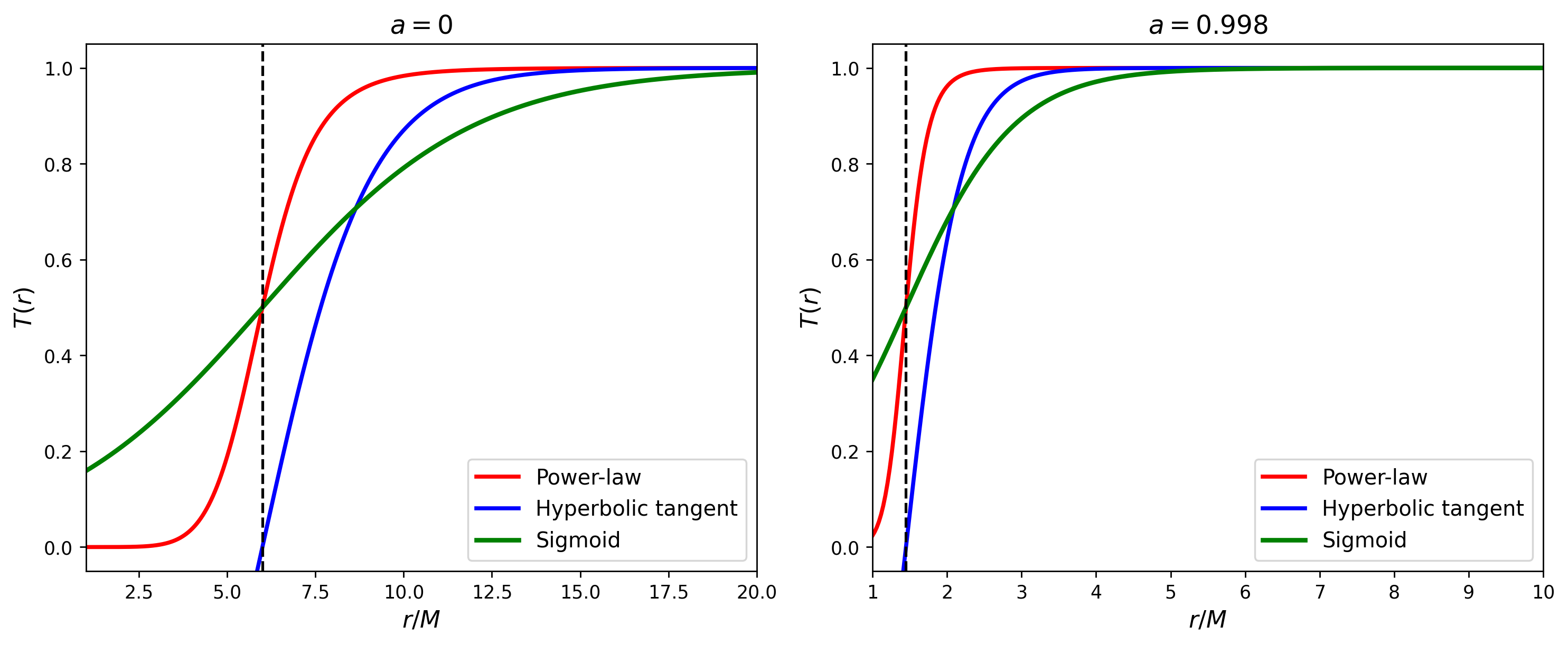}
	\caption{Comparison of transition functions $T(r)$ for two black hole spin parameters. 
		{Left panel:} $a=0$ with $r_{\rm ISCO}=6M$. 
		{Right panel:} $a=0.998$ with $r_{\rm ISCO}\approx1.24M$. 
		The logistic sigmoid (green), power-law (red, p=0.6), and hyperbolic tangent (blue) are shown for each case. 
		The vertical dashed line marks the ISCO radius.}
	\label{fig:transition}
\end{figure}

Physically, viscous and MHD stresses maintain nearly Keplerian rotation outside the ISCO \cite{hawley2001global}. As the gas approaches the ISCO, centrifugal support gradually weakens, angular momentum transport becomes less effective, and the radial velocity increases toward the free-fall value \cite{potter2021full}. Therefore, the transition between these two regimes is expected to be continuous rather than abrupt. We seek a transition function $T(r)$ that smoothly connects these two regimes.  After testing several functional forms for a transition function, including \( T(r)=\frac{r^{p}}{r^{p}+r_{0}^{p}} \) (power-law) and \( T(r)=\tanh\left(\frac{r-r_{0}}{wr_{0}}\right) \) (hyperbolic tangent), we adopt the logistic sigmoid because it provides a smooth, bounded transition with an adjustable width parameter and good numerical stability (Fig.~\ref{fig:transition}):

\begin{equation}
	T(r;r_{0},w)=
	\frac{1}{1+\exp\left[-\dfrac{r-r_{0}}{wr_{0}}\right]},
	\label{eq:sigmoid}
\end{equation}

where $w$ controls the width of the transition region. The black hole spin enters the model in two ways: (i) the location of the transition is determined by the spin-dependent ISCO radius, and (ii) the explicit expressions for the Keplerian and free-fall components (Table \ref{tab:keplerian_freefall}) depend on the spin parameter $a$. The detailed structure of the transition function is summarized in Table~\ref{tab:transition_values}.

This framework is purely kinematic, interpolating between two regimes of test-particle geodesics without explicitly taking into account magnetic fields, associated pressure, flux, or stress tensors. Since the governing equations lack explicit magnetic physics, the underlying mathematical form applies broadly to any sub-Keplerian RIAF, whether in a MAD or standard and normal evolution (SANE) state. The MAD cases reflect our choice of calibration target rather than the internal physics of the model; the three free parameters ($\alpha$, $\beta$, and $w$) are tuned specifically to fit the time- and azimuthally averaged radial profiles from the MAD-state GRMHD simulations of \cite{narayan2022jets}. In this intermediate (sub-Keplerian) regime, the radial four-velocity and angular velocity are given by a weighted blend of the nearly Keplerian and free-fall solutions:
\begin{equation}
	u^{r}_{\rm sub} = u^{r}_{\rm K} + \bigl[1-T(r)\cdot\alpha\bigr] \bigl(u^{r}_{\rm ff}-u^{r}_{\rm K}\bigr),
	\label{eq:sub_blend1}
\end{equation}
\begin{equation}
	\Omega_{\rm sub} = \Omega_{\rm K} + \bigl[1-T(r)\cdot \beta\bigr]  \bigl(\Omega_{\rm ff}-\Omega_{\rm K}\bigr).
	\label{eq:sub_blend2}
\end{equation}

\begin{figure}[h!]
	\centering
	\includegraphics[width=1.01\textwidth]{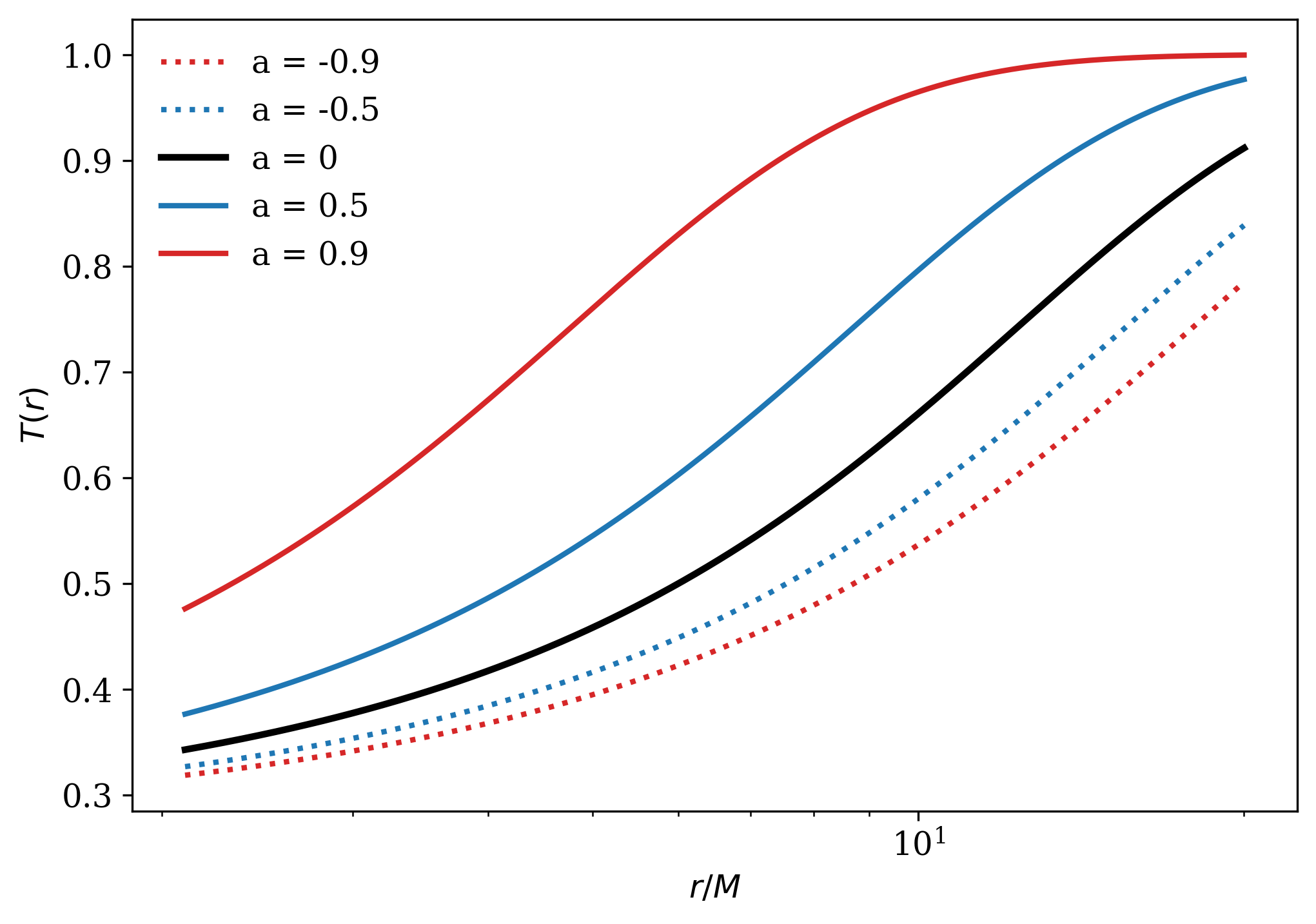}
	\caption{Radial dependence of the transition function $T(r)$ used in the flow model for $a=-0.9$, $a=-0.5$, $a=0$, $a=0.5$ and $a=0.9$. The function $T(r)$ is used to modulate the interpolation between the Keplerian and free-fall branches by varying the parameters $\alpha$ and $\beta$.}
	\label{fig:Transition}
\end{figure}

\begin{table}[H]
	\centering
	\caption{Radial variation of the transition function $T(r)$ evaluated at some selected radii for different spin parameters $a$. The transition radius is taken to be $r_0 = r_{\rm ISCO}$, corresponding to each black hole spin for retrograde and prograde accretion flows.}
	\label{tab:transition_values}
	\renewcommand{\arraystretch}{1.3}
	\begin{tabular}{cccccc}
		\toprule
		$r/M$ & $a=-0.9$ & $a=-0.5$ & $a=0.0$ & $a=0.5$ & $a=0.9$ \\
		\midrule
		2.5 & 0.329 & 0.339 & 0.358 & 0.399 & 0.519 \\
		6 & 0.423 & 0.449 & 0.500 & 0.603 & 0.830 \\
		10 & 0.537 & 0.580 & 0.661 & 0.796 & 0.965 \\
		20 & 0.785 & 0.839 & 0.912 & 0.976 & 1.000 \\
		\bottomrule
	\end{tabular}
\end{table}

\section{Model Fitting}

\begin{figure}[h]
	\centering
	\includegraphics[width=1.01\textwidth]{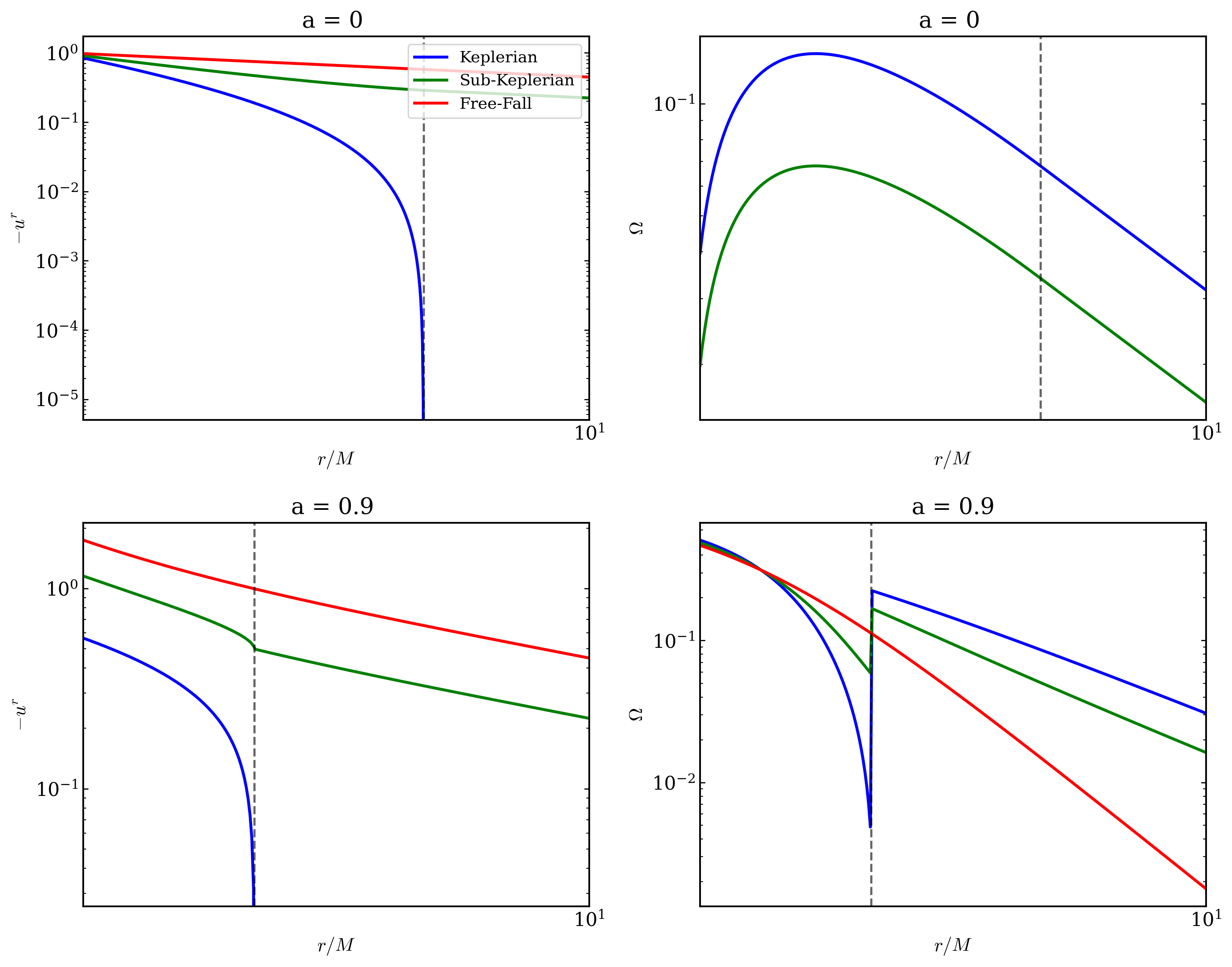}
	\caption{Comparison of radial velocity $-u^r$ and angular velocity $\Omega$ profiles for different accretion flow models around Schwarzschild ($a=0$, top panels) and rapidly rotating Kerr ($a=0.9$, bottom panels) black holes.}
	\label{fig:flow_dynamics_concise}
\end{figure}

Following \cite{Pu2016}, we first consider the constant-coefficient case $\alpha=\beta=0.5$, which represents a moderately sub-Keplerian flow. The corresponding profiles for black-hole spins $a=0$ and $a=0.9$ are shown in Figure \ref{fig:flow_dynamics_concise} based on Equations \ref{eq:sub_no_blend1} and \ref{eq:sub_no_blend2}. This formulation is general and can be applied to black holes with different spin parameters. Near the ISCO, turbulent velocity fluctuations allow fluid elements to move inward from finite radii, making the transition to the plunging region gradual rather than sharply defined \cite{mummery2024dynamics}. This behaviour is consistent with the two-component advective flow (TCAF) paradigm, in which a Keplerian disk coexists with a low angular momentum, sub-Keplerian halo \cite{ChakrabartiTitarchuk1995}.

\begin{figure}[h!]
	\centering
	\includegraphics[width=1.01\columnwidth]{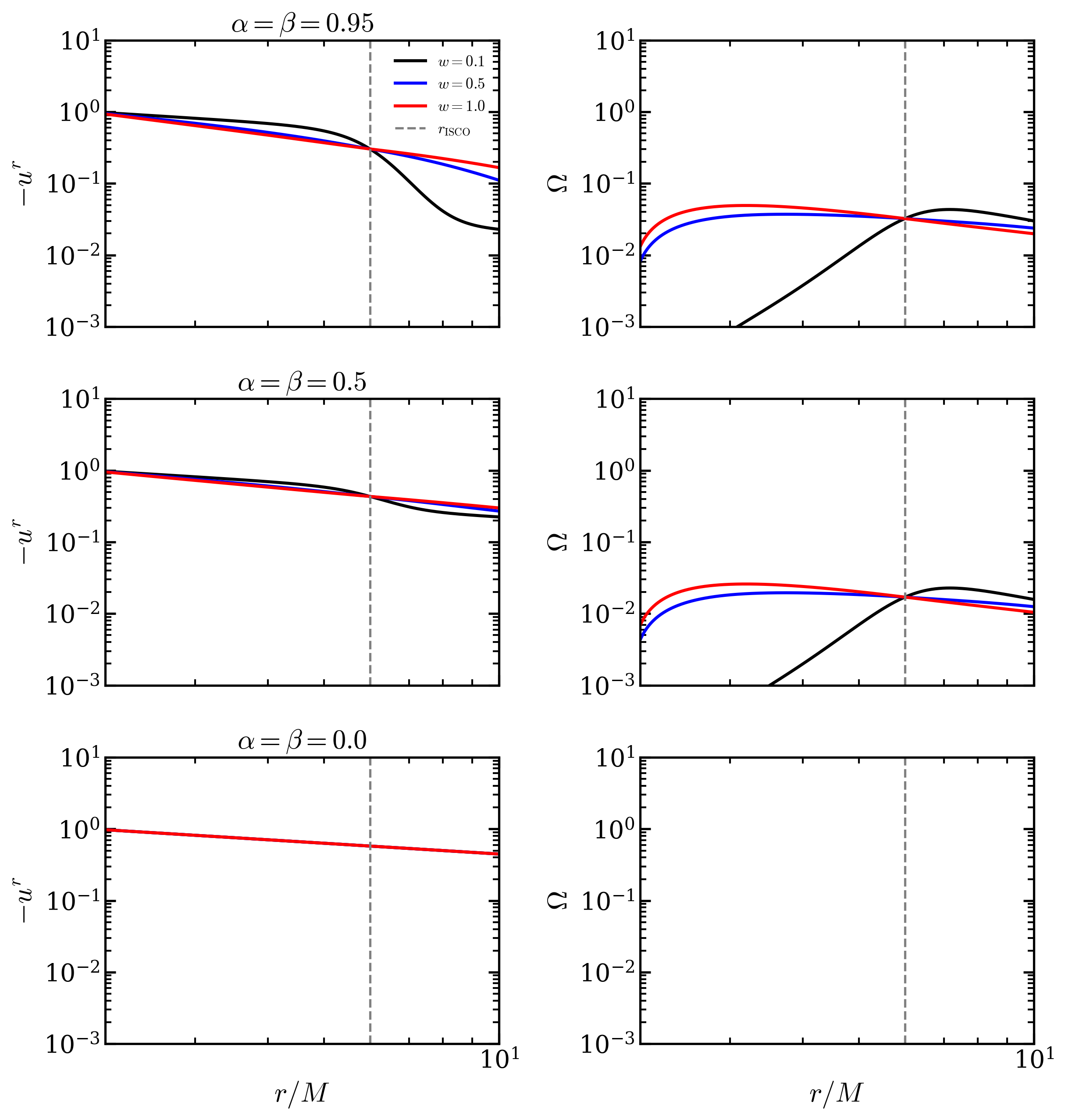}
	\caption{
		Radial velocity ($-u^r$, left panels) and angular velocity ($\Omega$, right panels) profiles for a non-rotating black hole ($a = 0$). Each row corresponds to $\alpha = \beta = 0.95$, $0.5$, and $0.0$ (top to bottom), while different colored curves represent transition widths $w = 0.1$, $0.5$, and $1.0$. The vertical dashed line indicates the ISCO radius. 
	}
	\label{fig:spin-}
\end{figure}

\begin{figure}[h!]
	\centering
	\includegraphics[width=\columnwidth]{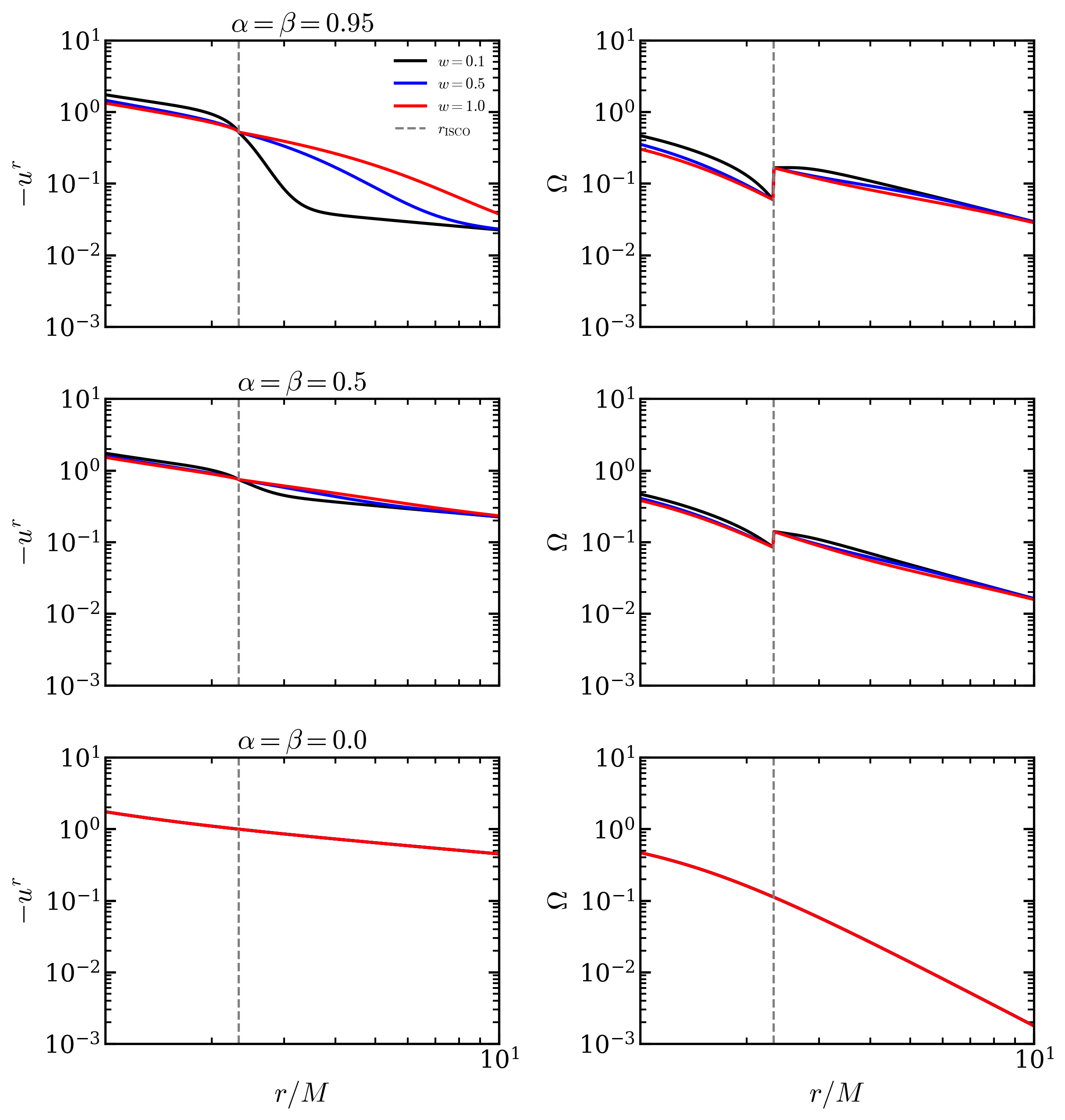}
	\caption{
		Radial velocity ($-u^r$, left panels) and angular velocity ($\Omega$, right panels) profiles for a rotating black hole with spin $a = 0.9$. The location of the ISCO shifts inward due to the black hole spin, and differences between flow models become more pronounced in the inner region.
	}
	\label{fig:spin+}
\end{figure}

Figures~\ref{fig:spin-} and~\ref{fig:spin+} show the radial velocity ($-u^r$, left panels) and angular velocity ($\Omega$, right panels) profiles for non-rotating ($a=0$) and rapidly rotating ($a=0.9$) black holes, obtained using Equations~(\ref{eq:sub_blend1}) and~(\ref{eq:sub_blend2}). The rows correspond to $\alpha=\beta=0.95$, $0.5$, and $0.0$ (top to bottom), while the colored curves represent transition widths $w=0.1$, $0.5$, and $1.0$. The adopted parameter set $(\alpha,\beta,w)$ reproduces the physically expected near Keplerian rotation at large radii and a smooth transition to free fall close to the black hole, and was selected by minimizing the root-mean-square error (RMSE) across all five black hole spin cases with respect to the numerical velocity profiles of \cite{narayan2022jets}.

\begin{table}[H]
	\centering
	\footnotesize
	\caption{Radial velocity ($-u^{r}$) and angular velocity ($\Omega$) of the Keplerian, sub-Keplerian, and free-fall models at selected radii for the  Schwarzschild black hole, $a=0$.}
	\label{tab:fitting_values}
	\begin{tabular}{lccc}
		\toprule
		\textbf{Flow Model} & \textbf{$r/M=3$} & \textbf{$r/M=6$} & \textbf{$r/M=10$} \\
		\midrule
		
		Keplerian &
		\begin{tabular}{@{}c@{}}
			$-u^r=0.33$\\
			$\Omega=0.14$
		\end{tabular} &
		\begin{tabular}{@{}c@{}}
			$-u^r=0.00$\\
			$\Omega=0.07$
		\end{tabular} &
		\begin{tabular}{@{}c@{}}
			$-u^r=0.00$\\
			$\Omega=0.03$
		\end{tabular} \\
		\midrule
		Sub-Keplerian &
		\begin{tabular}{@{}c@{}}
			$-u^r=0.57$\\
			$\Omega=0.07$
		\end{tabular} &
		\begin{tabular}{@{}c@{}}
			$-u^r=0.30$\\
			$\Omega=0.03$
		\end{tabular} &
		\begin{tabular}{@{}c@{}}
			$-u^r=0.22$\\
			$\Omega=0.01$
		\end{tabular} \\
		\midrule
		Free-fall &
		\begin{tabular}{@{}c@{}}
			$-u^r=0.82$\\
			$\Omega=0.00$
		\end{tabular} &
		\begin{tabular}{@{}c@{}}
			$-u^r=0.58$\\
			$\Omega=0.00$
		\end{tabular} &
		\begin{tabular}{@{}c@{}}
			$-u^r=0.45$\\
			$\Omega=0.00$
		\end{tabular} \\
		
		\bottomrule
	\end{tabular}
\end{table}

\begin{table}[ht]
	\centering
	\footnotesize
	\caption{Average  RMSE between the semi-analytical and numerical radial velocity profiles as a function of the transition width $w$, with $\alpha=\beta=0.95$. The numerical profiles are taken from \cite{narayan2022jets}.}
	\label{tab:w_rmse}
	\begin{tabular}{ccc}
		\toprule
		\textbf{$w$} & Average  RMSE & Remarks \\
		\midrule
		0.1 & 0.6807 & Sharp transition; may be too abrupt \\
		0.5 & 0.4502 & Moderate transition width \\
		1.0 & 0.1507 & Standard choice; smooth transition \\
		\bottomrule
	\end{tabular}
\end{table}	

For $\alpha=\beta=0.95$, the flow remains nearly Keplerian at large radii, whereas decreasing $\alpha$ and $\beta$ produces progressively more sub-Keplerian flows (see Figure~\ref{fig:flow_dynamics_concise}). Representative velocity profiles for the Schwarzschild case ($a=0$) are listed in Table~\ref{tab:fitting_values}. At $r/M=10$, the fitted values ($-u^r=0.22$, $\Omega=0.01$) remain sub-Keplerian, while at $r/M=6$ ($-u^r=0.30$, $\Omega=0.03$) the flow is distinctly sub-Keplerian. Closer to the black hole, at $r/M=3$, the values ($-u^r=0.57$, $\Omega=0.07$) indicate approaching free fall. Applying the same fitting procedure to the Kerr case ($a=0.9$) yields a similarly smooth transition between the Keplerian, sub-Keplerian, and free-fall regimes. Among the models considered, $\alpha=\beta=0.95$ with $w=1.0$ provides the best agreement with the numerical simulations of hot, thick MAD accretion flows (see Table \ref{tab:w_rmse}). Furthermore, the presented transition model is calibrated only over the radial interval covered by the GRMHD simulations. Beyond the simulation computational boundary, the flow is assumed to asymptotically approach the Keplerian solution. Therefore, the analytical transition function is not intended to describe the flow outside the simulated domain at arbitrarily large radius.

\section{Results and Discussion}

 In this section, we compare the semi-analytical model with the time- and azimuthally averaged GRMHD simulation results of \cite{narayan2022jets}.

\subsection{Radial velocity}

	\begin{figure}
	\centering
	
	\begin{subfigure}{0.495\textwidth}
		\centering
		\includegraphics[width=\textwidth]{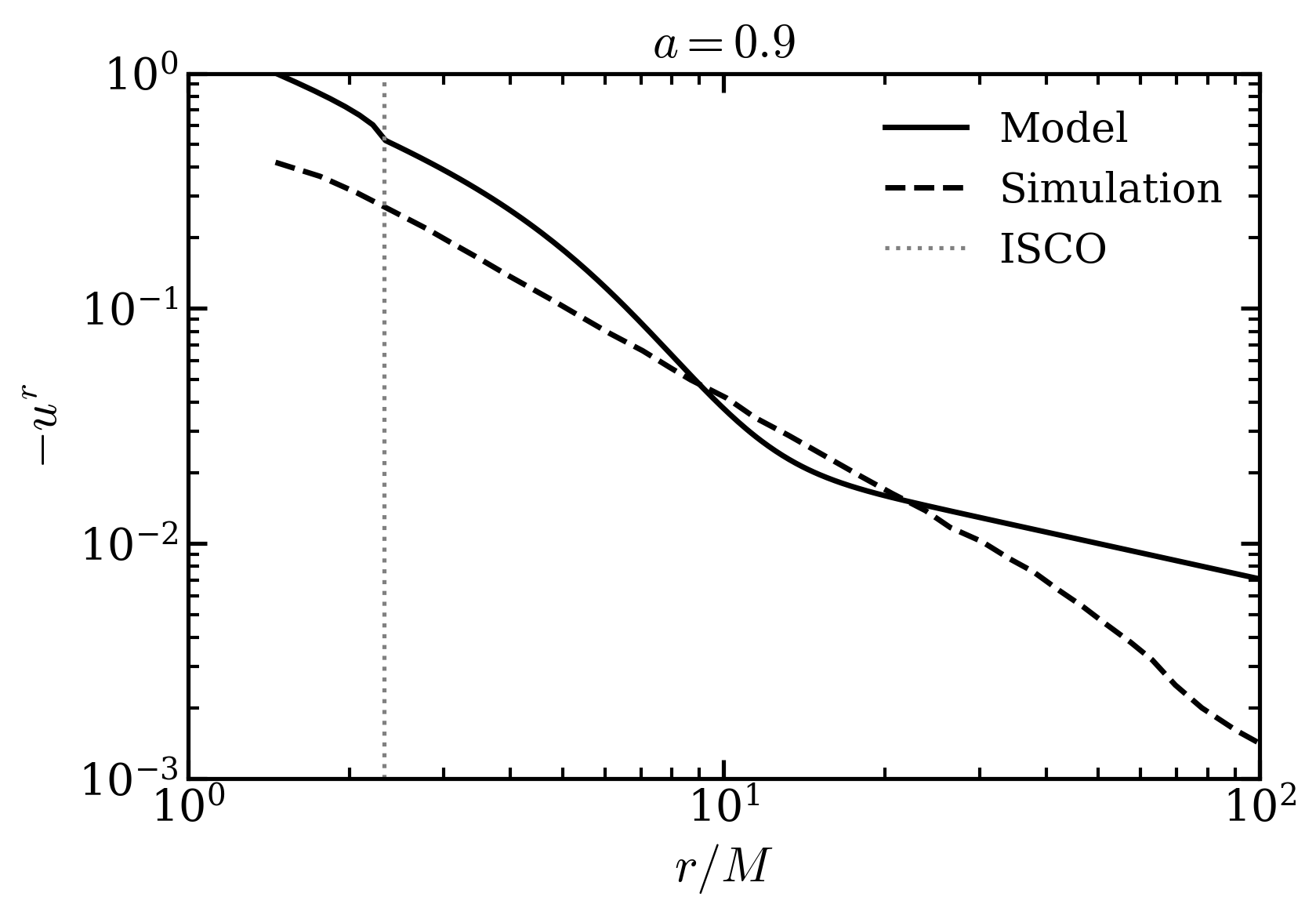}
		\label{fig:spin00}
	\end{subfigure}
	\hfill
	\begin{subfigure}{0.495\textwidth}
		\centering
		\includegraphics[width=\textwidth]{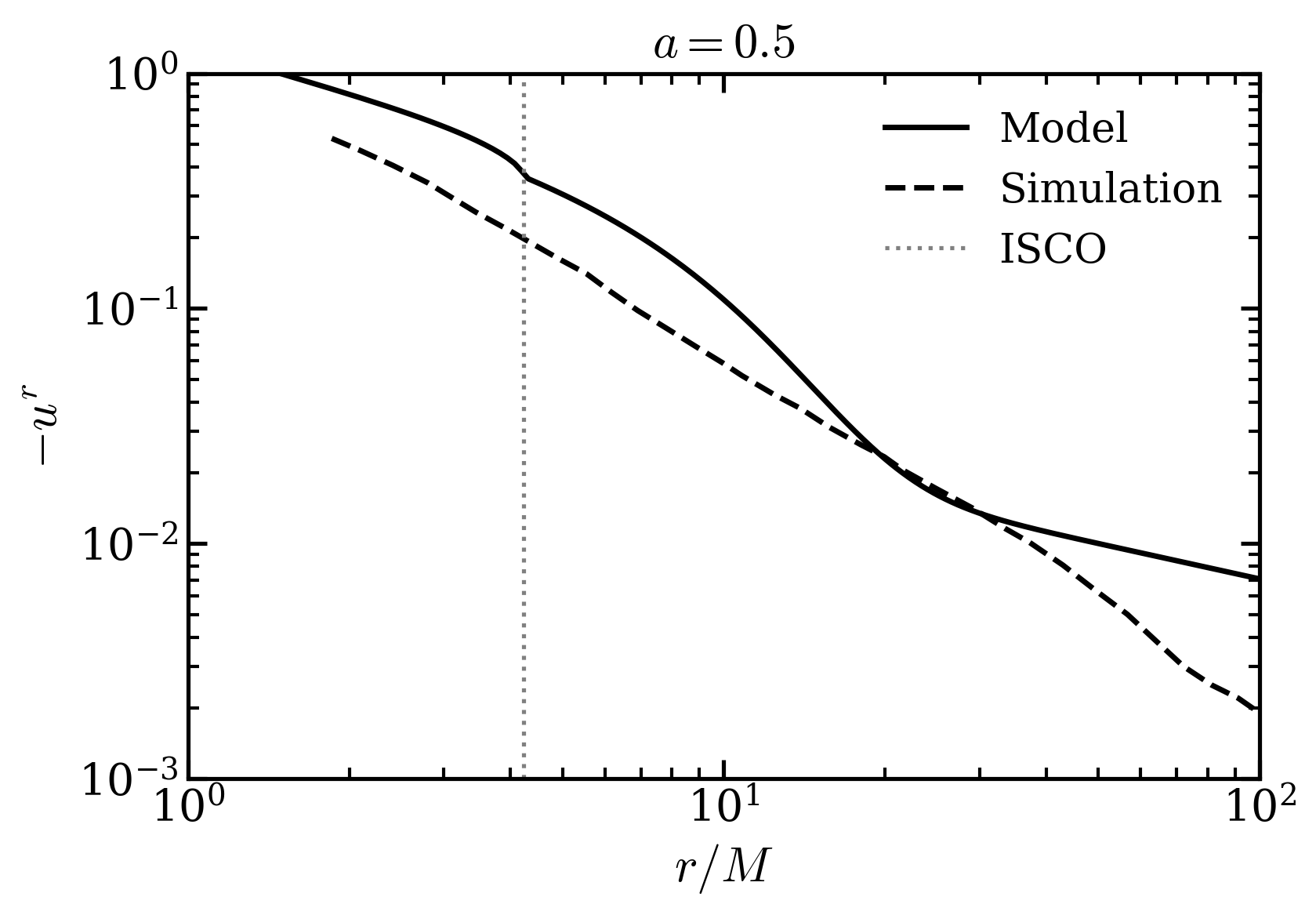}
		\label{fig:spin05}
	\end{subfigure}
	
	\vspace{0.5cm}
	
	\begin{subfigure}{0.495\textwidth}
		\centering
		\includegraphics[width=\textwidth]{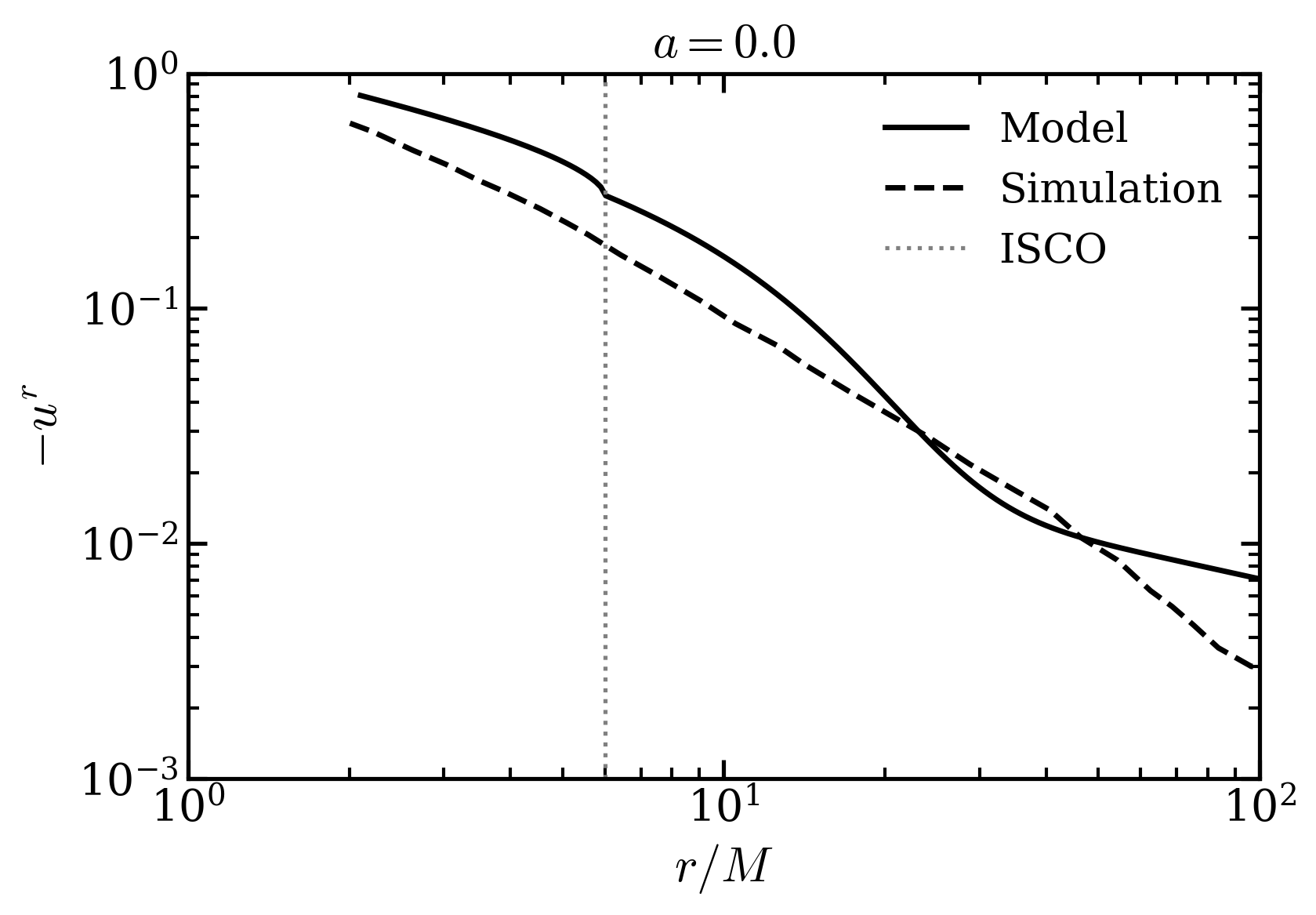}
		\label{fig:spin09}
	\end{subfigure}
	\hfill
	\begin{subfigure}{0.495\textwidth}
		\centering
		\includegraphics[width=\textwidth]{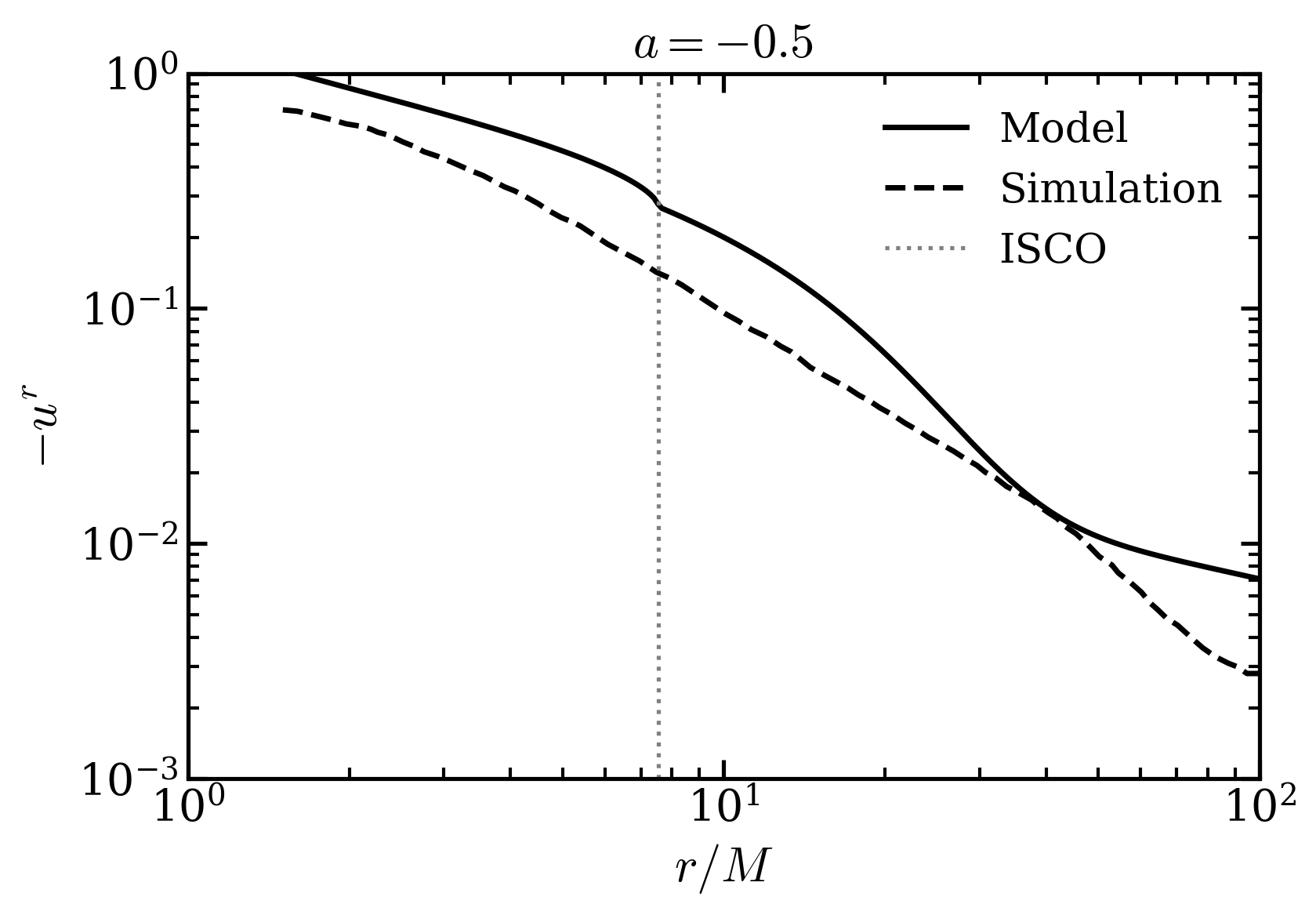}
		\label{fig:aneg09}
	\end{subfigure}
	
	\vspace{0.5cm}
	
	\begin{subfigure}{0.495\textwidth}
		\centering
		\includegraphics[width=\textwidth]{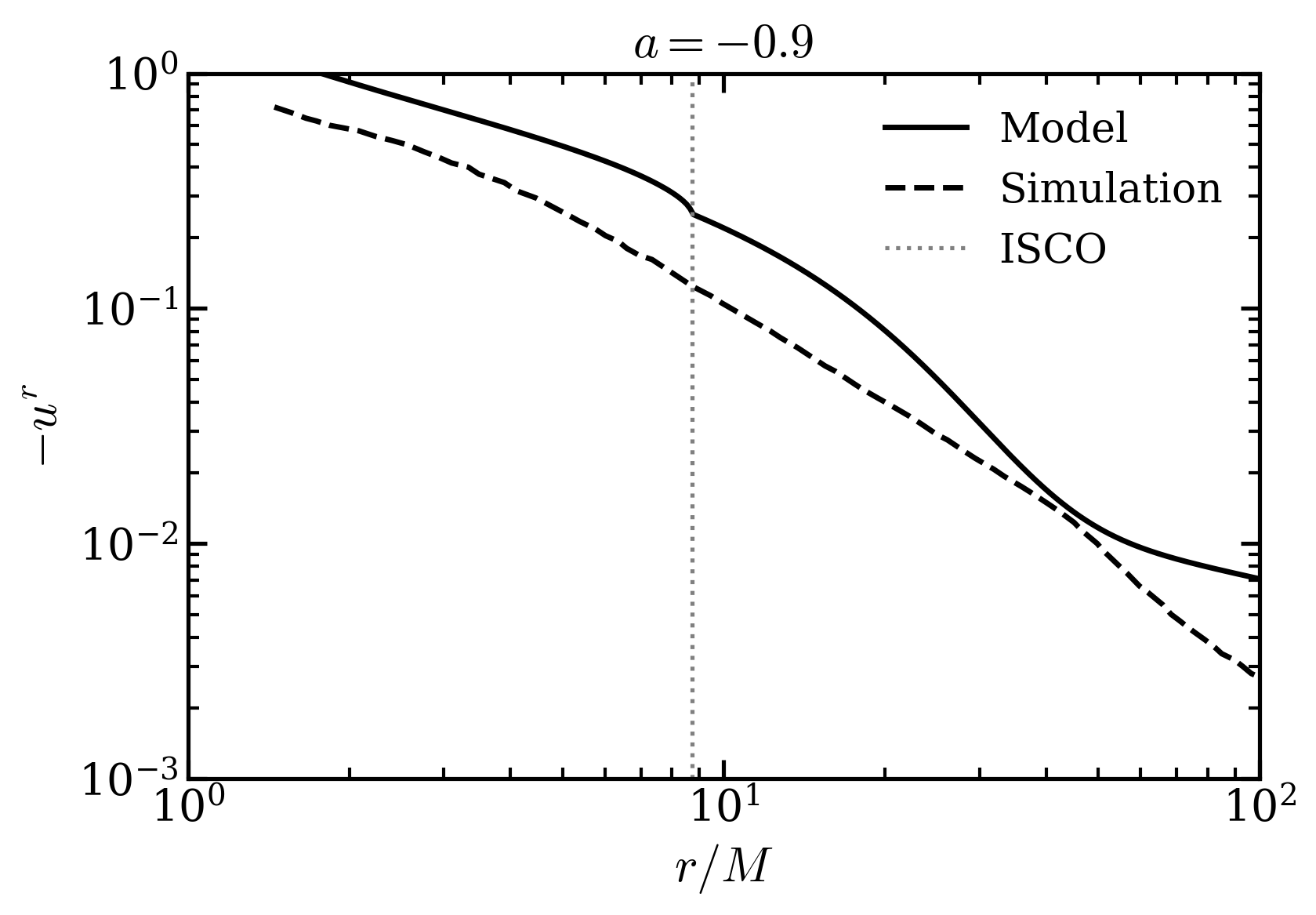}
		\label{fig:aneg05}
	\end{subfigure}
	
	\caption{
		Radial velocity profiles ($-u^r$) as a function of radius ($r/M$) for different black hole spin parameters ($a = -0.9, -0.5, 0.0, 0.5, 0.9$), showing comparison between the present theoretical model (solid lines) and simulation results (dashed lines). Noticeable differences persist in the inner accretion flow region.
	}
	\label{fig:radial_velocity_all}
\end{figure}

	The model proposed by \cite{Pu2016} shows partial agreement with the simulation results of \cite{narayan2022jets} in reproducing the velocity profile. However, by incorporating a transition function to account for the gradual dynamical changes in the flow, the resulting velocity profiles show improved agreement with time-averaged profiles from the GRMHD simulations of RIAFs~\cite{narayan2022jets}.

	\begin{table}[t]
		\centering
		\footnotesize
		\caption{Error between the theoretical model and numerical simulation for the radial velocity $-u^r$.}
		\label{tab:velocity_agreement}
		\begin{tabular}{cccc}
			\toprule
			Spin $a$ & MAE & RMSE & Avg.\ Factor \\
			\midrule
			$-0.9$ & 0.1264 & 0.1811 & 1.808 \\
			$-0.5$ & 0.1067 & 0.1487 & 1.719 \\
			$0.0$  & 0.0812 & 0.1210 & 1.572 \\
			$+0.5$ & 0.0887 & 0.1427 & 1.831 \\
			$+0.9$ & 0.0771 & 0.1602 & 1.956 \\
			\midrule
			Average & 0.0960 & 0.1507 & 1.777 \\
			\bottomrule
		\end{tabular}
	\end{table}

	In ADAFs, the inward radial velocity is not negligible, unlike in standard thin disk models. Instead, it becomes a significant fraction of the free-fall velocity \cite{takahashi2007advection}. This characteristic is preserved in our model. The present results are in qualitative agreement with the RIAF solutions of \cite{manmoto2000advection}, which show that black hole spin strongly influences the radial infall velocity. The numerical simulations of \cite{narayan2022jets} show that prograde flows generally exhibit smaller radial velocities than retrograde ones, a feature successfully captured by our model (Figure~\ref{fig:radial_velocity_all}).  A notable feature of our radial velocity profiles is the absence of any clear signature of the ISCO, which is consistent with \cite{page1974disk}. The radial velocity approaches the speed of light near the event horizon for all spins. The average RMSE is $0.1507$, with a corresponding mean absolute error (MAE) of $0.096$, indicating deviations within a factor of $\approx 1.8$ (cf. Table \ref{tab:velocity_agreement}). The model successfully reproduces the radial trend of the inflow velocity across all spin values.

	\subsection{Angular Velocity}

	\begin{table}[t]
		\centering
		\caption{Error estimates between the theoretical model and numerical simulation for the angular velocity $\Omega$.}
		\label{tab:omega_agreement}
		\begin{tabular}{cccc}
			\toprule
			Spin $a$ & MAE & RMSE & Avg.\ Factor \\
			\midrule
			$-0.9$ & 0.0331 & 0.0461 & 2.046 \\
			$-0.5$ & 0.0138 & 0.0225 & 1.733 \\
			$0.0$  & 0.0088 & 0.0113 & 1.927 \\
			$+0.5$ & 0.0227 & 0.0459 & 1.286 \\
			$+0.9$ & 0.0159 & 0.0309 & 1.229 \\
			\midrule
			Average & 0.0189 & 0.0314 & 1.644 \\
			\bottomrule
		\end{tabular}
	\end{table}

\begin{figure}[htbp]
    \centering
	
	\begin{subfigure}{0.495\textwidth}
		\centering
		\includegraphics[width=\textwidth]{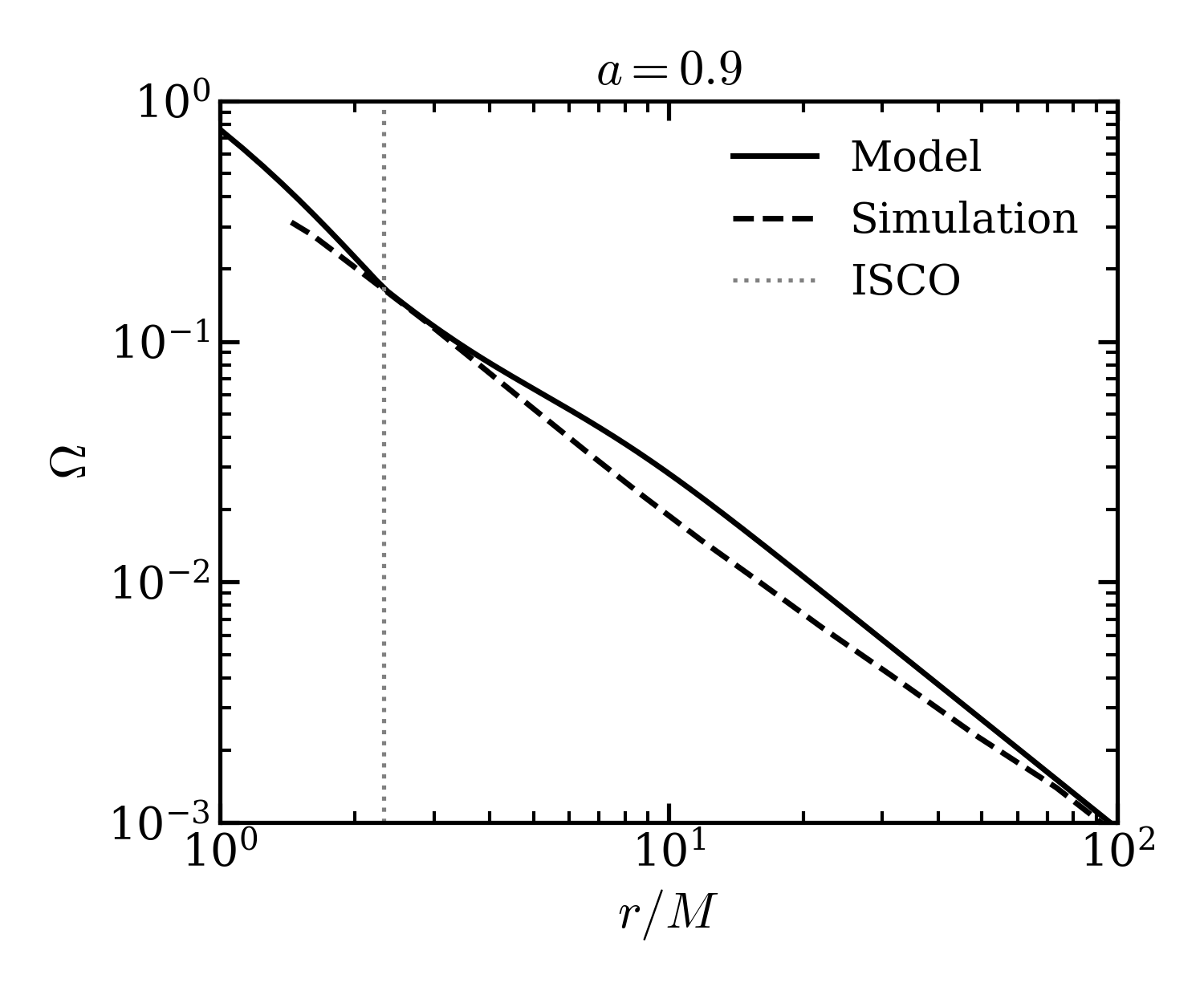}
		\label{fig:density00}
	\end{subfigure}
	\hfill
	\begin{subfigure}{0.495\textwidth}
		\centering
		\includegraphics[width=\textwidth]{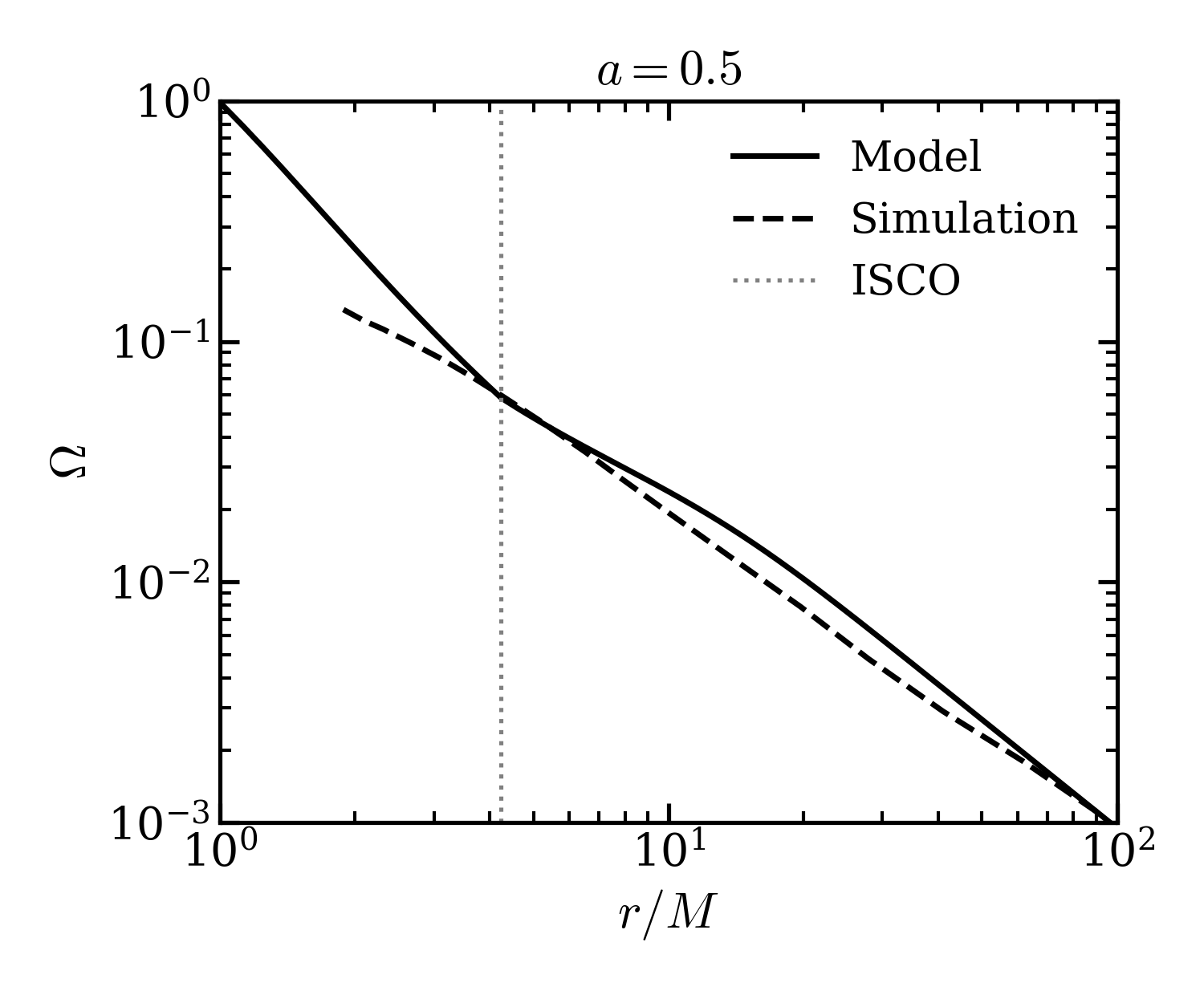}
		\label{fig:omega05}
	\end{subfigure}

	\begin{subfigure}{0.495\textwidth}
		\centering
		\includegraphics[width=\textwidth]{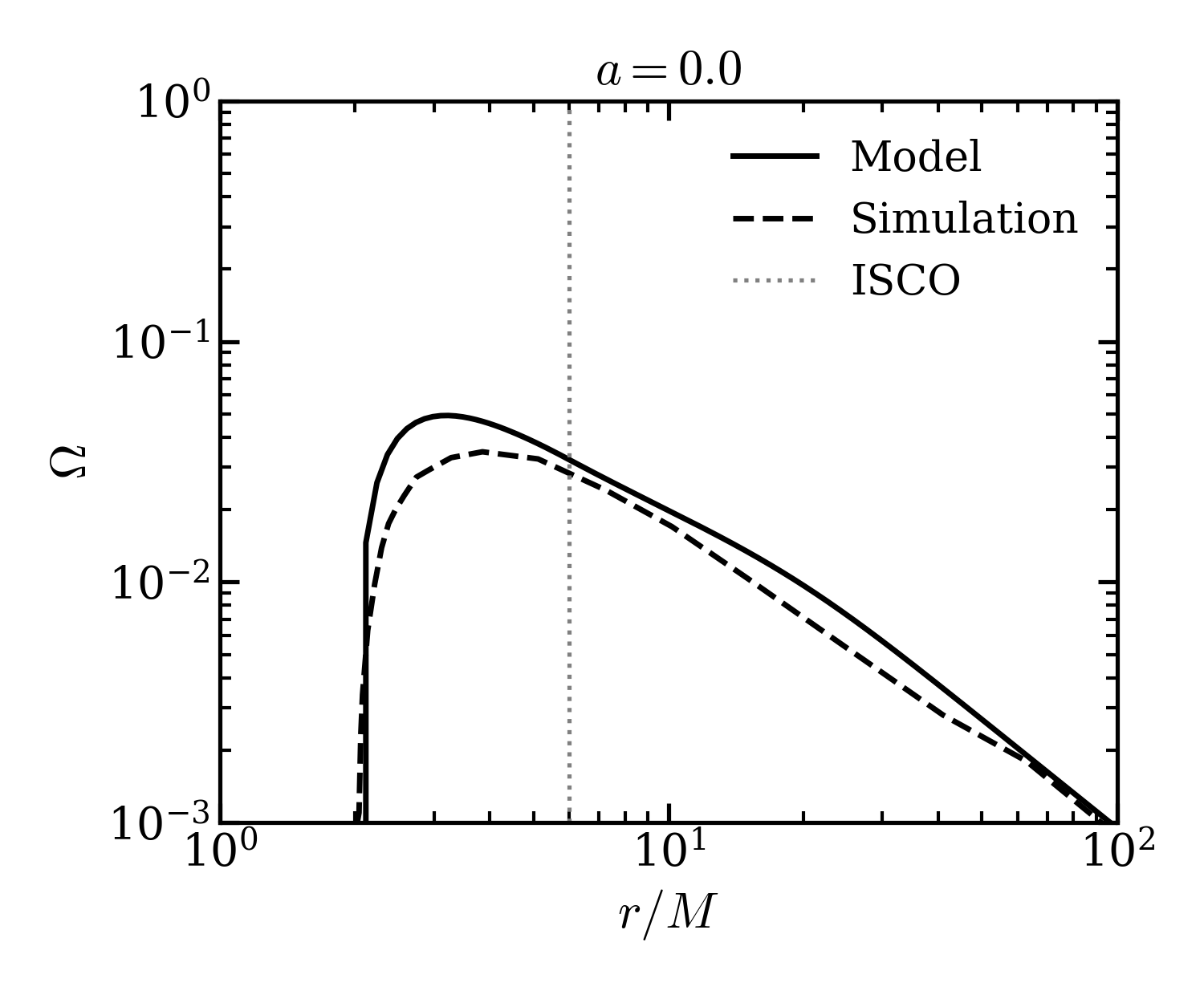}
		\label{fig:omega09}
	\end{subfigure}
	\hfill
	\begin{subfigure}{0.495\textwidth}
		\centering
		\includegraphics[width=\textwidth]{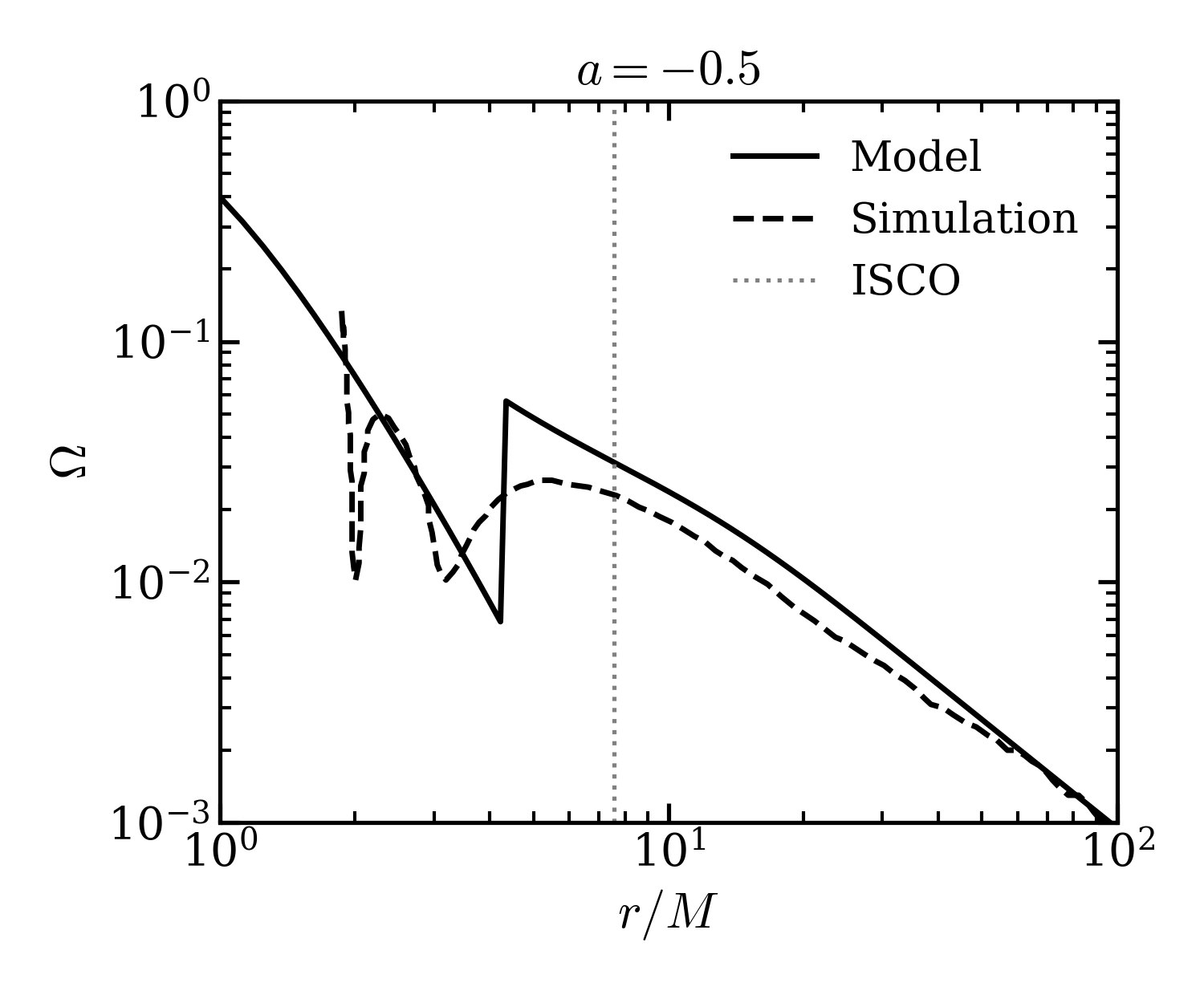}
		\label{fig:omega_n05}
	\end{subfigure}

	\begin{subfigure}{0.495\textwidth}
		\centering
		\includegraphics[width=\textwidth]{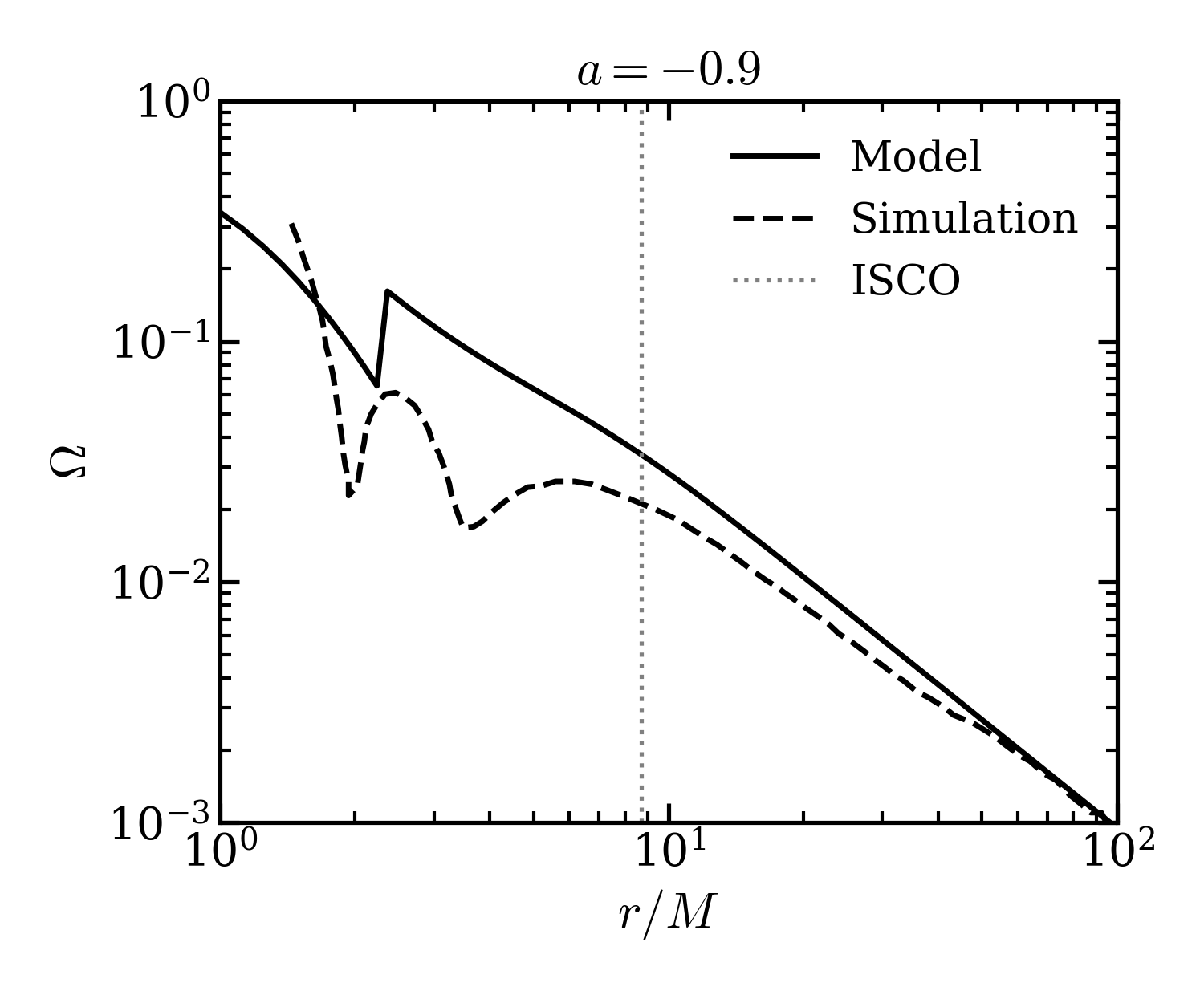}
		\label{fig:omega_n09}
	\end{subfigure}
	
	\caption{
		Angular velocity profiles ($\Omega$) as a function of radius ($r/M$) for different black hole spin parameters ($a = -0.9, -0.5, 0.0, 0.5, 0.9$), comparing the theoretical model (solid lines) with simulation results (dashed lines). The model employs a smooth transition function to connect the nearly Keplerian rotation outside the ISCO with the plunging region inside it. The radial trend of the angular velocity is well reproduced, although moderate deviations remain in the inner region. 
	}
	\label{fig:angular_velocity_all}
\end{figure}

 The sharp cusp in the retrograde angular-velocity profiles is an intrinsic feature of the blending function. It arises because the Keplerian and free-fall angular velocities have opposite radial gradients near the transition region for retrograde spins, producing a local extremum in the blended profile rather than a numerical artifact (Figure~\ref{fig:angular_velocity_all}).

At larger radii, the influence of black hole spin on the angular velocity becomes negligible, consistent with previous studies \cite{popham1998advection}. This behaviour is also evident in both the simulation results and our modelled profiles. The average RMSE is $0.0314$ for angular velocity, with a corresponding MAE of $0.0189$, indicating deviations within a factor of $\approx 1.6$ (cf. Table \ref{tab:omega_agreement}).

\subsection{Density}

	\begin{table}[t]
		\centering
		\caption{Error estimates between the theoretical model and numerical simulation for the density $\rho$. }
		\label{tab:density_agreement}
		\begin{tabular}{cccc}
			\toprule
			Spin $a$ & MAE & RMSE & Avg.\ Factor \\
			\midrule
			$-0.9$ & 0.0059 & 0.0104 & 1.423 \\
			$-0.5$ & 0.0035 & 0.0064 & 1.337 \\
			$0.0$  & 0.0019 & 0.0026 & 1.432 \\
			$+0.5$ & 0.0071 & 0.0094 & 1.728 \\
			$+0.9$ & 0.0080 & 0.0103 & 2.036 \\
			\midrule
			Average & 0.0053 & 0.0078 & 1.591 \\
			\bottomrule
		\end{tabular}
	\end{table}

\begin{figure}[htbp]
	\centering
	
	\begin{subfigure}{0.495\textwidth}
		\centering
		\includegraphics[width=\textwidth]{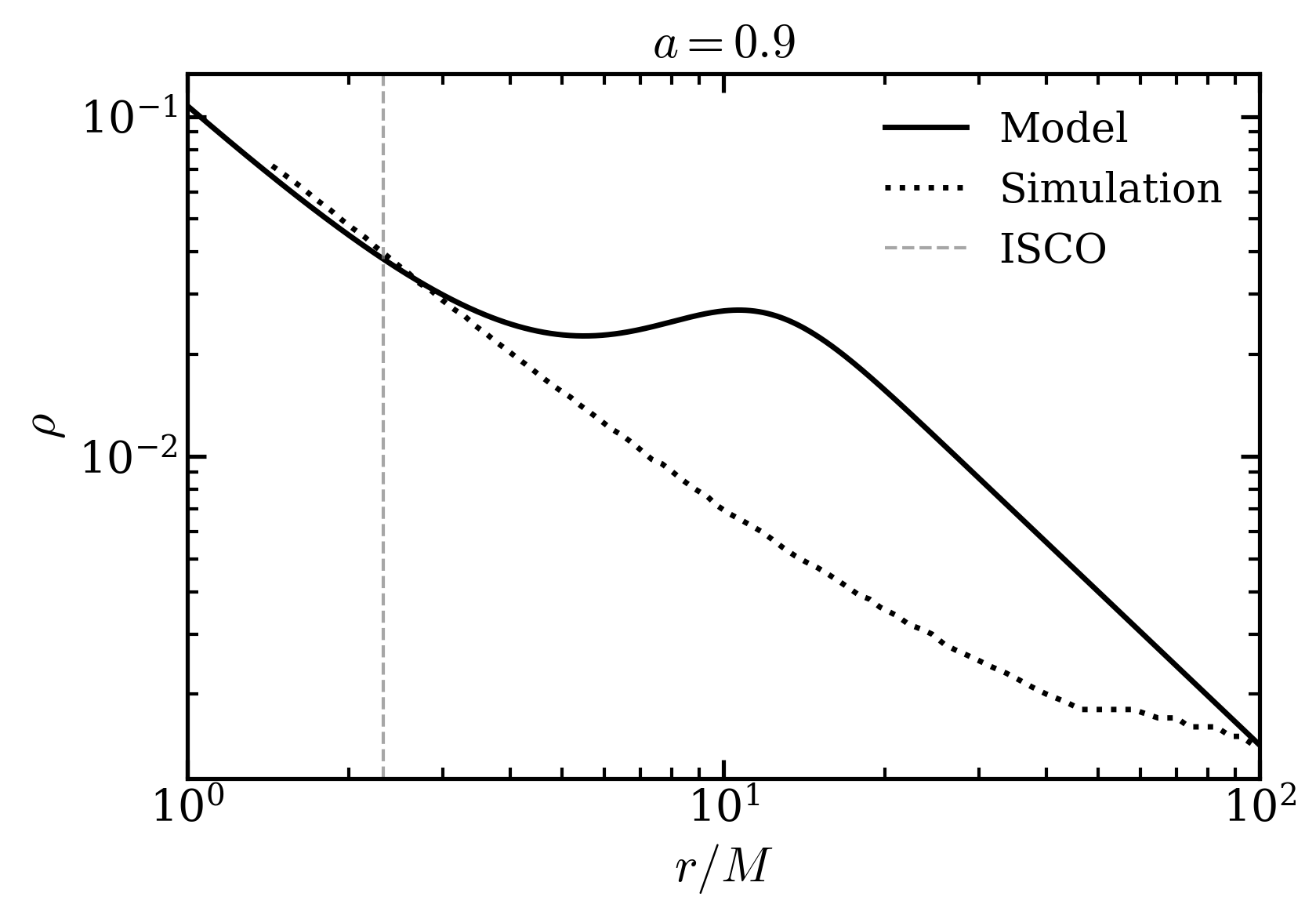}
		\label{fig:omega00}
	\end{subfigure}
	\hfill
	\begin{subfigure}{0.495\textwidth}
		\centering
		\includegraphics[width=\textwidth]{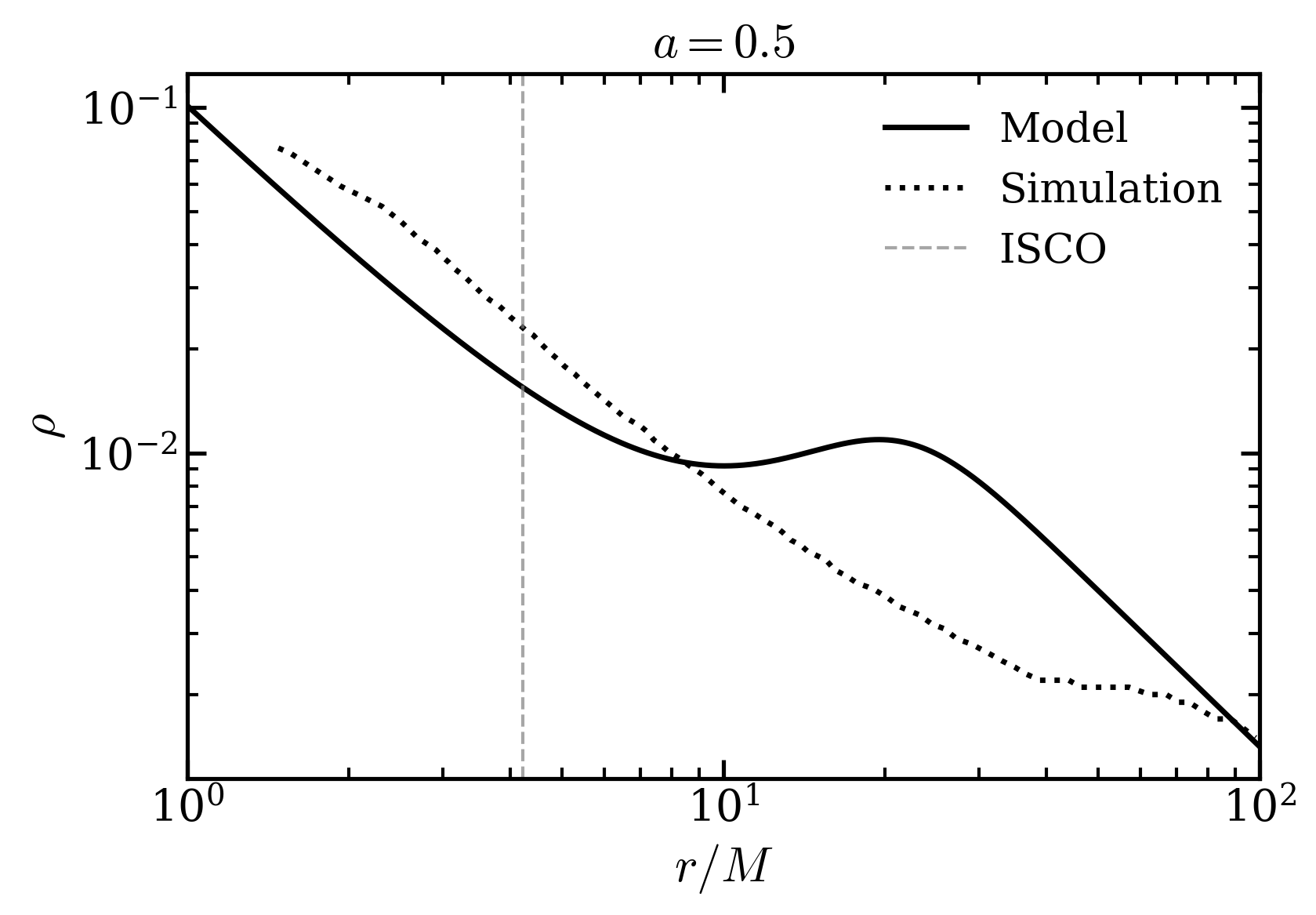}
		\label{fig:density05}
	\end{subfigure}

	\begin{subfigure}{0.495\textwidth}
		\centering
		\includegraphics[width=\textwidth]{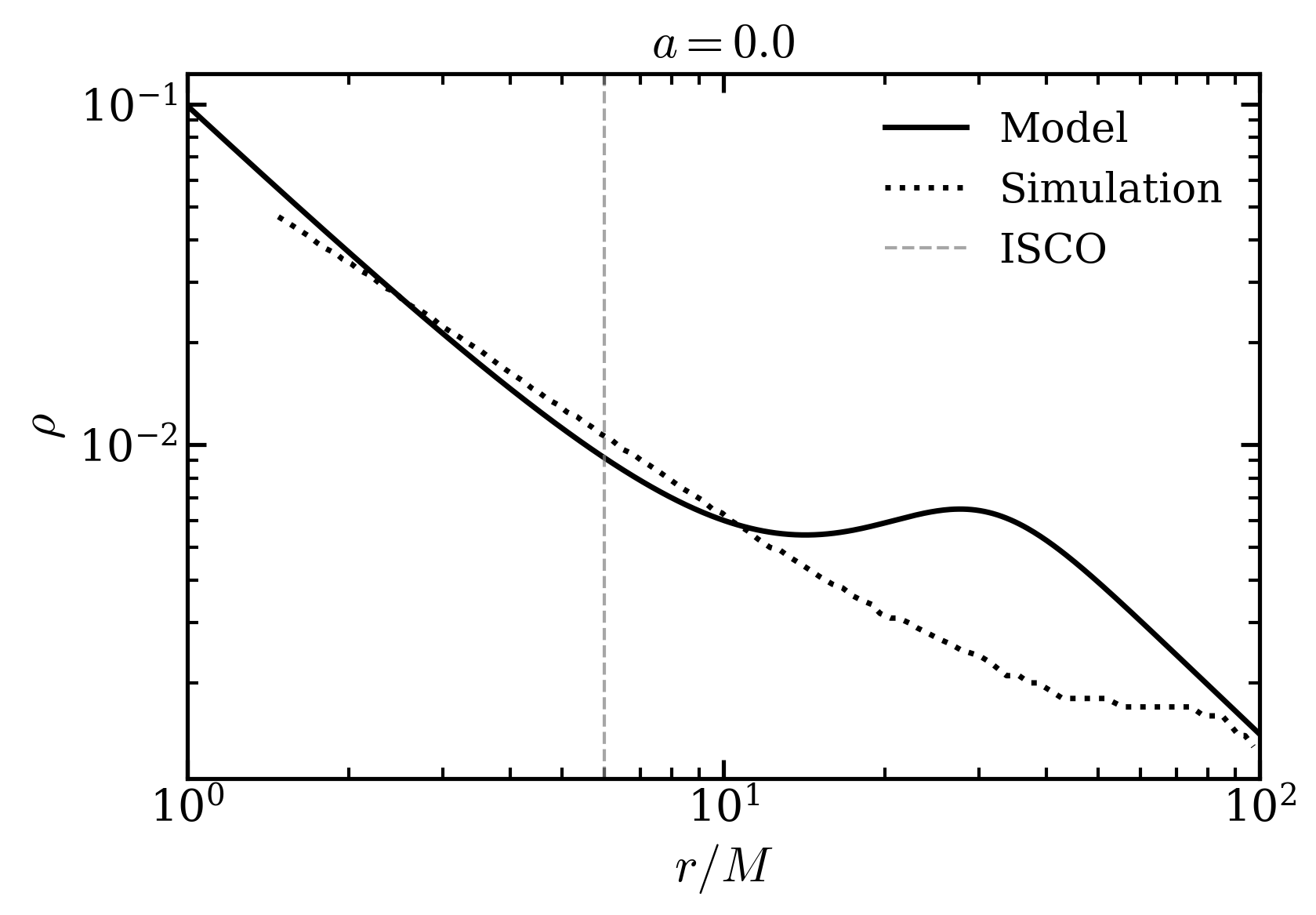}
		\label{fig:density09}
	\end{subfigure}
	\hfill
	\begin{subfigure}{0.495\textwidth}
		\centering
		\includegraphics[width=\textwidth]{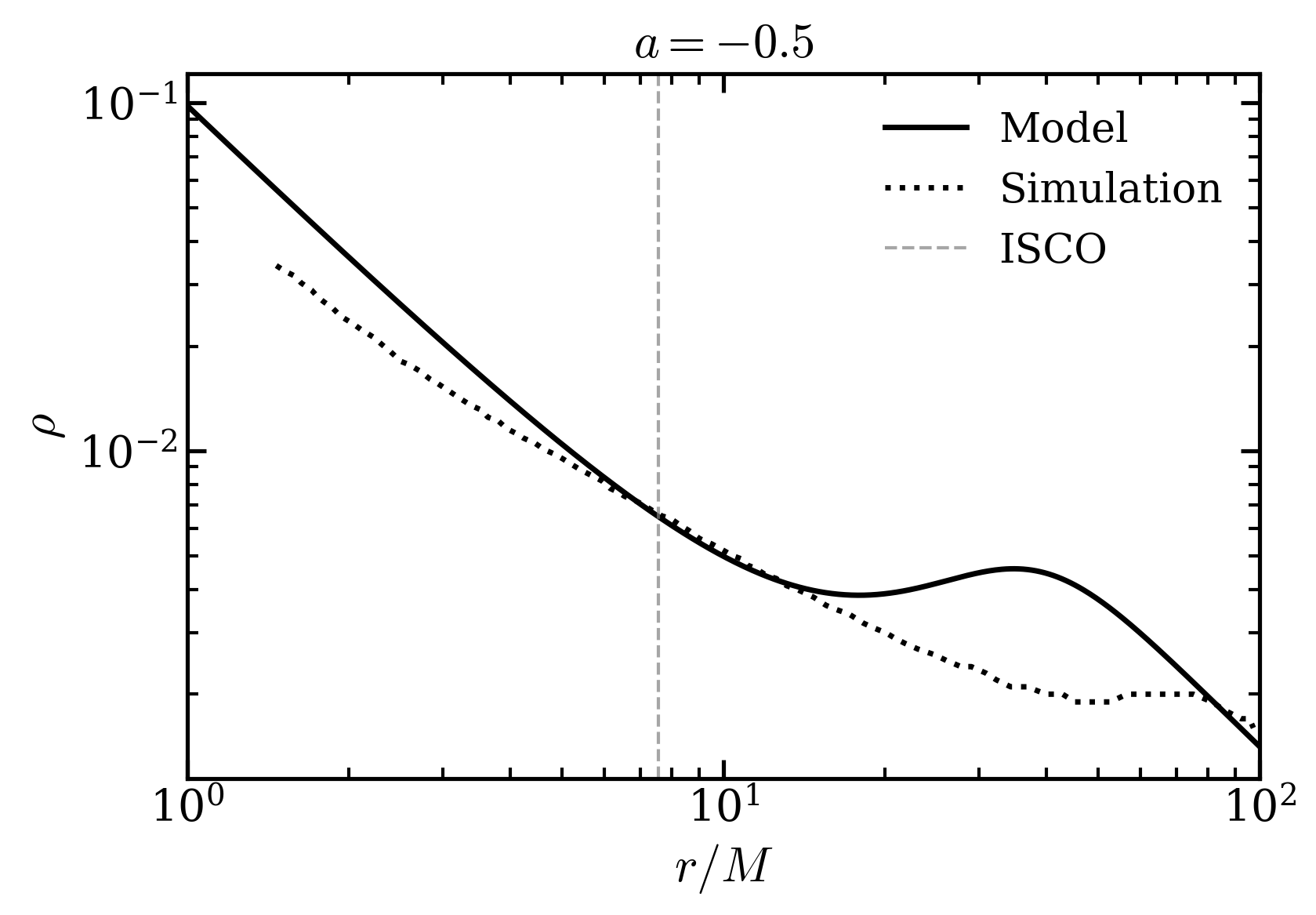}
		\label{fig:aneomega09}
	\end{subfigure}

	\begin{subfigure}{0.495\textwidth}
		\centering
		\includegraphics[width=\textwidth]{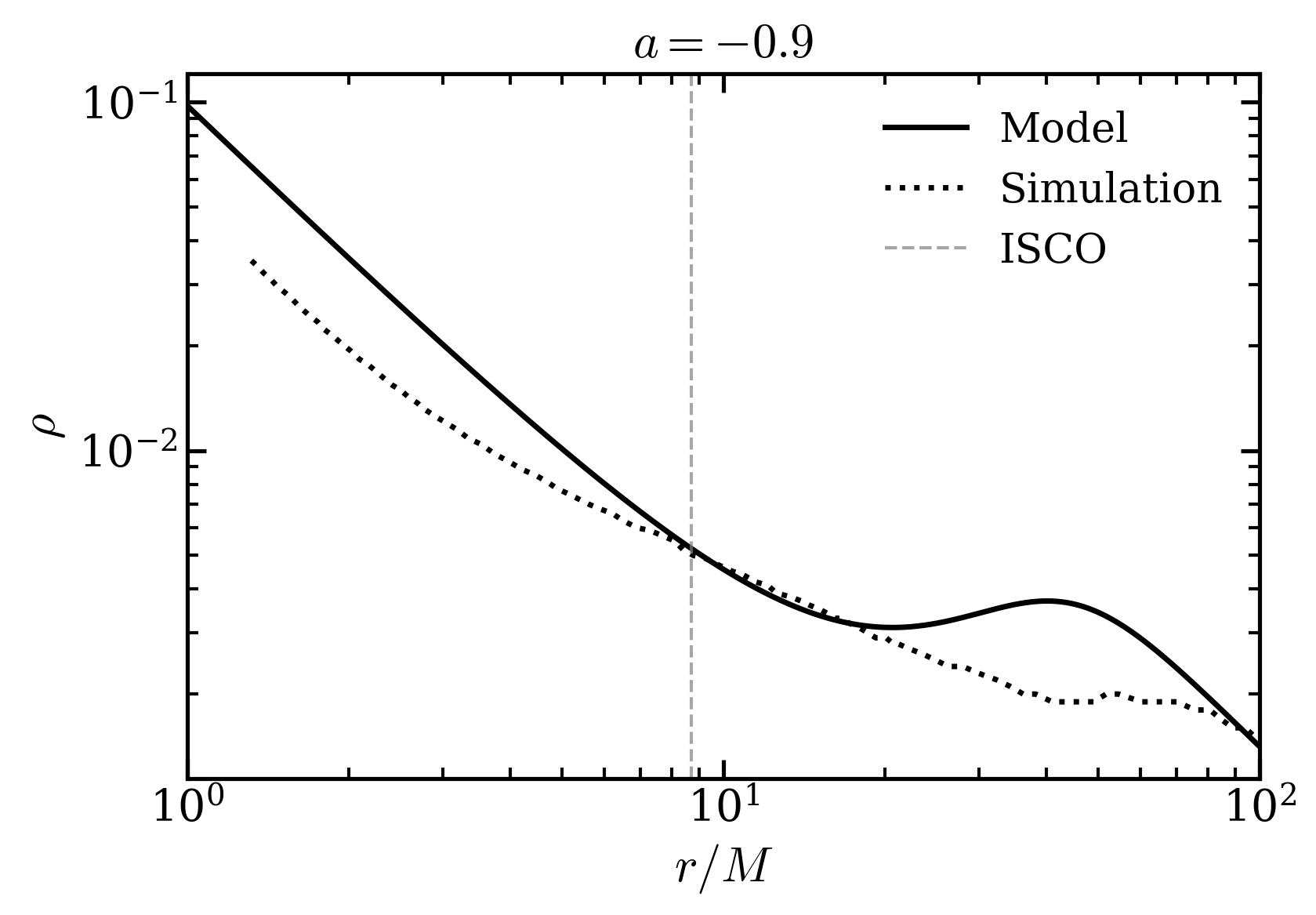}
		\label{fig:aneomega05}
	\end{subfigure}
	
	\caption{
		Density profiles as a function of radius ($r/M$) for different black hole spin parameters ($a = -0.9, -0.5, 0.0, 0.5, 0.9$), comparing the present theoretical model (solid lines) with simulation results (dashed lines).  The radial behaviour of the density is well reproduced across all spin values, although moderate differences persist in the inner region. 
	}
	\label{fig:density_all}
\end{figure}

From the simulation density profiles, prograde flows exhibit higher inner-region density coupled with slower radial infall; density remains relatively elevated in prograde flows compared to retrograde ones despite comparable mass accretion rates \cite{narayan2022jets}. No distinct break appears at the ISCO in the density distribution; the density structure is shaped predominantly by radial inflow dynamics (cf. Figure \ref{fig:density_all}). The density profiles peak near the equatorial plane. Contrary to the profile suggested by \citet{manmoto1997spectrum} , the density does not decrease as the gas flows inward from the outer boundary, except in the specific case of a Schwarzschild black hole. The agreement between the semi-analytical density model and the simulation data is quantified for both prograde and retrograde flows. The average RMSE for the density is 0.0078, with a corresponding MAE of 0.0053, indicating deviations within a factor of $\approx 1.6$. The semi-analytical model successfully captures the radial behaviour of the density profile across all spin values as shown in Table  \ref{tab:density_agreement}.

To verify that the radially varying blend of Equations ~(\ref{eq:sub_blend1}) and
(\ref{eq:sub_blend2}) introduces a genuine structural improvement over the constant-coefficient model of \cite{Pu2016} , rather than simply a better-tuned version of the
same functional freedom, Table~\ref{tab:alpha_beta_comparison} reports the
identical error metrics, computed using the same methodology as in the previous sections--for the
constant $\alpha=\beta=0.5$ baseline, evaluated against the same
\cite{narayan2022jets} profiles.

\begin{table}[t]
	\centering
\caption{Error estimates for different values of $\alpha$ and $\beta$. The left columns show the results obtained without the transition function ($\alpha=\beta=0.5$), while the right columns show the results obtained using the transition function with $\alpha=\beta=0.95$.}	\label{tab:alpha_beta_comparison}
	\begin{tabular}{l c c c c c c}
		\toprule
		& \multicolumn{3}{c}{$\alpha=\beta=0.5$} & \multicolumn{3}{c}{$\alpha=\beta=0.95$} \\
		\cmidrule(lr){2-4} \cmidrule(lr){5-7}
		Observable & MAE & RMSE & Avg.\ Factor & MAE & RMSE & Avg.\ Factor \\
		\midrule
		$-u^r$  & 0.1407 & 0.1502 & 8.306 & 0.0960 & 0.1507 & 1.777 \\
		$\Omega$   & 0.0219 & 0.0387 & 1.841  & 0.0189 & 0.0314 & 1.644 \\
		$\rho$  & 0.0059 & 0.0097 & 2.601  & 0.0053 & 0.0078 & 1.591 \\
		\bottomrule
	\end{tabular}
\end{table}

\subsection{Specific Angular Momentum}

	\begin{figure}[htbp]
	\centering
	\includegraphics[width=0.75\textwidth]{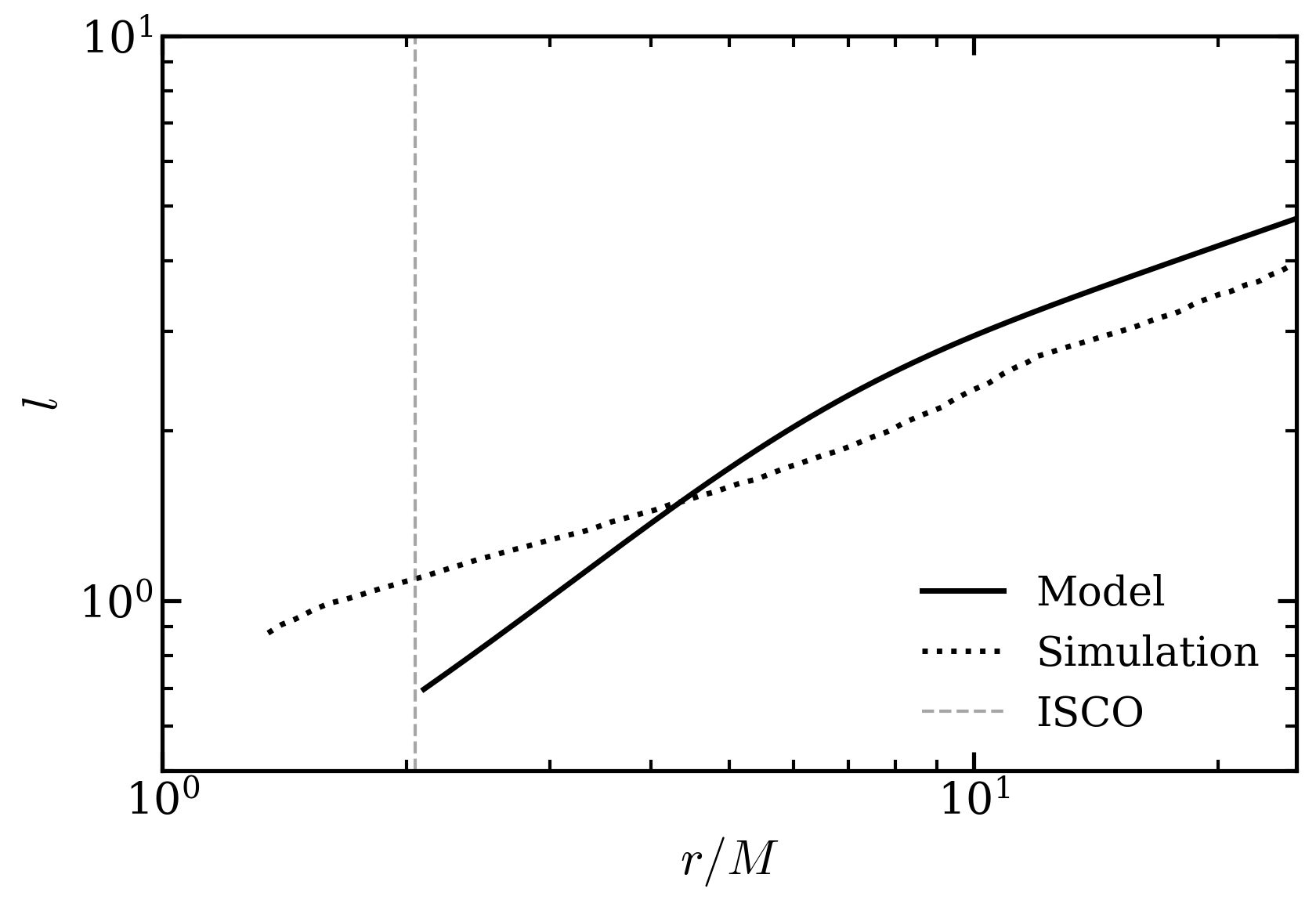}
	\caption{ Equatorial profile of the specific angular momentum, $\ell$, obtained from time- and azimuthally averaged simulation data (dotted curve) compared with the theoretical model prediction (solid curve), for a = 0.9375.
	}
	\label{fig:acc2}
\end{figure}

In Kerr spacetime, the specific angular momentum $\ell$ is a conserved quantity along a geodesic. The time- and azimuthally averaged specific angular momentum, $\ell$ is shown as the black dotted curve in Figure~\ref{fig:acc2}. As in previous GRMHD studies \cite{chael2021}, the simulated profile remains substantially sub-Keplerian at all radii. Whereas earlier analytical models approximated the simulation profile using simple power-law fits, the present model employs a smooth transition between the nearly Keplerian and free-fall branches. The resulting profile reproduces the radial behaviour of the simulation, with $\mathrm{RMSE} = 0.5324$ and $\mathrm{MAE} = 0.4743$, corresponding to agreement within a factor of $\approx 1.2$. 
Figure~\ref{fig:acc2} indicates that the model systematically underestimates the specific angular momentum in the innermost region. This discrepancy is a direct mathematical consequence of the chosen free-fall branch, which assumes zero angular momentum by construction, meaning $\ell_{\rm ff}\equiv0$ across all radii. In contrast, the simulated GRMHD flow approximately conserves its angular momentum past the ISCO, carrying that value through the plunging region.

\section{Conclusion}

We have developed a simple semi-analytical framework for the time-averaged structure of hot sub-Keplerian radiatively inefficient accretion flows around Kerr black holes, in which the transition from nearly Keplerian rotation to free-fall infall is governed by the radially varying blend weight defined in Equations~(\ref{eq:sub_blend1}) and (\ref{eq:sub_blend2}), with the transition function $T(r)$ tied to the spin-dependent ISCO radius. Comparisons with the long-duration, time-averaged GRMHD simulation data of \citet{narayan2022jets} show that the model reproduces the principal radial trends of the flow structure over a range of black hole spins. The model agrees with the simulation profiles to within factors of approximately $1.8$, $1.6$, and $1.6$ for the radial velocity, the angular velocity, and the density, respectively. The radial velocity profiles reproduce the systematic trend seen in the simulations, in which the prograde flows exhibit slower infall than retrograde ones. The absence of a sharp feature at the ISCO in both the modelled and simulated radial velocity profiles reflects the dominant role of radial inflow dynamics and magnetic support in the inner region, consistent with previous findings. The specific angular momentum profile exhibits sub-Keplerian values at all radii, with good quantitative agreement (factor $\approx 1.2$), indicating that the transition prescription successfully captures much of the angular momentum distribution, though not all of it reported in numerical simulations \cite{chael2021}.

Our approach captures the gradual loss of centrifugal support as matter approaches the black hole--a key improvement over earlier models that used constant parameters~\cite{Pu2016}. The success of the transition function demonstrates that the complex, time-averaged structure of GRMHD accretion flows can be captured by a remarkably simple low-dimensional description. This suggests that much of the essential radial structure of hot accretion flows is governed by a small number of effective degrees of freedom, making the model a fast, practical alternative to computationally expensive GRMHD simulations while retaining the principal physical behaviour. The goal of this study was to provide a simple yet physically motivated analytical model that reproduces the main features of state-of-the-art GRMHD simulations while remaining sufficiently transparent for future applications. In the future, this framework can readily be adapted in ray-tracing calculations of black hole shadows and spectra, as improved initial and boundary conditions for numerical simulations, or as a useful tool for exploring the parameter space of RIAFs.  

We also highlight a few limitations in the current framework. The model is strictly kinematic rather than dynamical; it does not explicitly solve for magnetic fields and associated parameters, meaning that any magnetic support is only implicitly encoded through calibration against the GRMHD data. The free-fall branch enforces zero angular momentum by construction, whereas the simulated plunging flow approximately conserves $\ell_{ISCO}$—a discrepancy that likely drives the near-horizon deviations in our angular velocity and specific angular momentum profiles. The transition function enforces a smooth, continuous radial profile that does not replicate the small-radius break visible in the simulations. Introducing an additional structural degree of freedom to capture this inner-region inflection could noticeably improve the model's accuracy. Addressing these limitations in future studies will provide a more robust basis for comparison with the baseline framework of \citet{Pu2016}.

\section*{Data Availability Statement}

This manuscript has no associated data.

\section*{Code Availability}

The codes used to generate the results in this manuscript are available from the corresponding author upon reasonable request.

\end{document}